\documentstyle[tighten,preprint,amsfonts,epsf,%
eqsecnum,aps,prd]{revtex}
\begin{document}
\draft

\hyphenation{
mani-fold
mani-folds
geo-metry
geo-met-ric
}

%% MACROS

%% uncomment the first two lines below if
%% amsfonts is not available
%%
%\def\Bbb{\bf}
%\def\frak{\bf}

\def\half{{\frac{1}{2}}}
\def\casehalf{{\case{1}{2}}}

\def\R{{\Bbb R}}
\def\BbbR{{\Bbb R}}
\def\BbbZ{{\Bbb Z}}
\def\BbbC{{\Bbb C}}

\def\RPthree{{{\Bbb RP}^3}}
\def\RPtwo{{{\Bbb RP}^2}}
\def\Z{{{\Bbb Z}}}

\def\AdSthree{\hbox{AdS${}_3$}}
\def\CAdSthree{\hbox{CAdS${}_3$}}
\def\Otwotwo{{\rm O}(2,2)}
\def\Dint{D_{\rm int}}

\def\Uone{{\rm U}(1)}

\def\Jcompact{J_c}

\def\ads{{Anti-de~Sitter}}

\def\arccosh{\mathop{\rm arccosh}\nolimits}

%% BEGINS

\narrowtext

\preprint{\vbox{\baselineskip=12pt
\rightline{SU-GP-98/8-1}
\rightline{hep-th/9808081}}}
\title{Single-exterior black holes and the AdS-CFT conjecture}
\author{Jorma Louko\footnote{%
Electronic address:
louko@aei-potsdam.mpg.de.
Address after September~1, 1999: 
School of Mathematical Sciences, 
University of Nottingham, 
Nottingham NG7 2RD, U.K\null. 
}}
\address{
Max-Planck-Institut f\"ur Gravitations\-physik,
Schlaatzweg~1,
D--14473 Potsdam,
Germany}
\author{Donald Marolf\footnote{Electronic address:
marolf@suhep.phy.syr.edu}}
\address{
Department of Physics,
Syracuse University,
Syracuse, New York 13244--1130, USA}
\date{Phys.\ Rev.\ D {\bf 59}, 066002 (1999)}
\maketitle
\begin{abstract}%
  In the context of the conjectured AdS-CFT correspondence of string
  theory, we consider a class of asymptotically \ads\ black holes
  whose conformal boundary consists of a {\em single\/} connected
  component, identical to the conformal boundary of \ads\ space. In a
  simplified model of the boundary theory, we find that the boundary
  state to which the black hole corresponds is pure, but this state
  involves correlations that produce thermal expectation values at the
  usual Hawking temperature for suitably restricted classes of
  operators. The energy of the state is finite and agrees in the
  semiclassical limit with the black hole mass. We discuss the
  relationship between the black hole topology and the correlations in
  the boundary state, and speculate on generalizations of the results
  beyond the simplified model theory.
\end{abstract}

\section{Introduction}

Black holes and related classical solutions
are a topic of long-standing interest in
string theory \cite{HS,Dark}.  Their study has shed light on
old questions \cite{bekenstein1,bekenstein2,Hawk} in black
hole physics 
(see e.g.\ \cite{SV}) 
as well as
dualities \cite{G,GGPT,AS}
and other stringy issues \cite{BKOP,RPS}.  
Indeed, it was an investigation of black holes that first lead to
Maldacena's conjecture \cite{MAL} 
(based on earlier work, e.g.\ \cite{GUKL}) 
relating string theory in asymptotically
Anti-de Sitter space to a conformal field theory on the boundary at
spatial infinity.  For evidence supporting this conjecture, see
\cite{evidence}. 

It is therefore natural to investigate asymptotically Anti-de Sitter
(AdS) black holes in light of Maldacena's conjecture.  Previous 
work \cite{MAST,BBG,TL,BDHM,BKLT,keskivak} 
has analyzed the (2+1)-dimensional
BTZ black holes \cite{BTZ} in this way,
using the fact that the classical black hole solutions
are certain quotients of AdS${}_3$ to identify associated states in the
conformal field theory (CFT).  Recall, however, that the 
nonextremal
BTZ black holes have two asymptotically Anti-de Sitter regions. 
As a result, the conformal boundary of such spacetimes is
not the usual $S^1 \times \R$ of the universal cover of
AdS${}_3$, but two copies
of this cylinder.   This means that, strictly speaking, 
such black holes are not described by quite the same conformal
field theory as AdS${}_3$ and
the corresponding states do not lie in the same Hilbert space.
We note that the $M=0$ black hole also has only a single asymptotic
region and so again cannot lie in the same Hilbert space as the BTZ
black holes.

In contrast, there are other asymptotically AdS black holes
which have only a single asymptotic region.  Some examples 
were constructed in \cite{ABBHP,ABH} as
quotients of AdS${}_3$. 
For such black holes, 
we expect the state in the boundary CFT
to be approximately described
by the result of a quotient-like operation on the
original vacuum $|0 \rangle$ of the conformal field theory.  We
do not consider here any effects which may result from additional
winding modes in the quotient spacetime.

Now, the
boundary state corresponding to the BTZ 
black hole has been characterized
as thermal \cite{MAST}.  This is a result of 
the boundary CFT having two disconnected components and the fact
that such black holes correspond to states of the boundary CFT
in which the two boundary components are entangled.  Thus, 
the boundary states are not pure
states on either boundary component separately.
On the other hand,
as discussed in \cite{horo-marolf-strc}, there is no reason to expect 
single-exterior black holes to be mixed states in any corresponding
sense.  In particular, 
as the boundary theory is now exactly the same as that of either
AdS${}_3$ or the
$M=0$ black hole, one expects to be able to interpret 
single-exterior
black holes as (pure state) excitations of these ground states.

In this paper we investigate the boundary states for certain
single-exterior, asymptotically AdS${}_3$ black holes by using a
simplified model of the boundary conformal field theory.  Our main
focus is on a class of spacetimes referred to as $\RPtwo$ geons, which
are analogous to the asymptotically flat $\RPthree$ geon
\cite{Nico,topocen,chamb-gibb,louko-marolf-geon}.  In
section \ref{classic} we discuss the structure and construction of the
$\RPtwo$ geons as quotient spaces of AdS${}_3$.  In section
\ref{quantum} we first motivate and define our model CFT and then
verify that the $\RPtwo$ geon corresponds to a pure state of our model
theory.  We also verify that the expectation value of the CFT
Hamiltonian in this state coincides with the mass of the black hole in
the limit where the black hole horizon circumference is much greater
than the length scale associated with the AdS space.  This corresponds
to the limit where $n_R$ or $n_L$ is much greater than $Q_1 Q_5$ in
terms of the left and right momentum, onebrane, and fivebrane quantum
numbers of the associated \cite{MAL} 
six-dimensional black string.
Note that taking such a limit is also important to remove quantum
corrections to the entropy of such black strings.  

After developing our technology
in the context of the $\RPtwo$ geons, we then briefly address 
the single-exterior black holes of Refs.\ 
\cite{ABBHP,ABH}
in section~\ref{Swedish}.
We close with some comments on the extrapolation of our results to the
full CFT of Maldacena's conjecture, the encoding of the geon topology
in the boundary state, and other issues in section \ref{disc}.

Our attention will be focused on 
the special case of Maldacena's conjecture
for the spacetime AdS${}_3 \times S^3 \times T^4$.
We use units in which $\hbar = c=1$. The 
(2+1)-dimensional Newton's constant is denoted by~$G_3$.

\section{$\AdSthree$, the spinless nonextremal BTZ hole, 
and the $\RPtwo$ geon}
\label{classic}

In this section we describe the quotient constructions of the spinless
nonextremal BTZ hole and the $\RPtwo$ geon from three-dimensional
\ads\ space, and the extension of this quotient construction to the
conformal boundaries of the spacetimes. The material for the BTZ hole
is familiar 
\cite{MAST,BTZ,horo-marolf-strc,carlip-rev}, 
but a review is needed in order
to establish the relationship between the BTZ hole and the geon. We
also mention generalizations of the geon construction to spacetimes
with additional internal dimensions, in particular the internal factor
$S^3 \times T^4$ that arises in string theory \cite{MAST}.

\subsection{$\AdSthree$, $\CAdSthree$, 
and the conformal boundary}

Recall that the three-dimensional \ads\ space $\AdSthree$ can be
defined as the surface
\begin{equation}
- l^2 = - {(T^1)}^2 - {(T^2)}^2 
+ {(X^1)}^2 + {(X^2)}^2 
\label{embedding-surface}
\end{equation}
in $\BbbR^{2,2}$ with the global coordinates 
$(T^1, T^2, X^1, X^2)$ and the metric 
\begin{equation}
ds^2 = 
- {(dT^1)}^2 - {(dT^2)}^2 
+ {(dX^1)}^2 + {(dX^2)}^2 
\ \ .
\label{embedding-metric}
\end{equation}
The positive parameter $l$ is the inverse of the Gaussian curvature.
$\AdSthree$ is a smooth three-dimensional spacetime with signature
$({-}{+}{+})$. It is maximally symmetric, and 
(the connected component of) the isometry group is 
(the connected component of) $\Otwotwo$. 
{}From now on we set $l=1$. 

It is useful to introduce on $\AdSthree$ the coordinates
$(t,\rho,\theta)$ by \cite{ABBHP}
\begin{mathletters}
\label{sausage-coords}
\begin{eqnarray}
T^1 &=&
\frac{1+\rho^2}{1-\rho^2} \cos t
\ \ ,
\\
T^2 &=&
\frac{1+\rho^2}{1-\rho^2} \sin t
\ \ ,
\\
X^1 &=&
\frac{2\rho}{1-\rho^2} \cos \theta
\ \ ,
\\
X^2 &=&
\frac{2\rho}{1-\rho^2} \sin \theta
\ \ .
\end{eqnarray}
\end{mathletters}
With $0\le \rho<1$ and 
the identifications 
\begin{equation}
(t,\rho,\theta) \sim 
(t,\rho,\theta+2\pi)
\sim 
(t+2\pi,\rho,\theta)
\ \ ,
\label{sausage-idents}
\end{equation}
these coordinates can be understood as global on $\AdSthree$, apart
from the elementary coordinate singularity at $\rho=0$. The metric
reads
\begin{equation}
ds^2 = \frac{4}{{(1-\rho^2)}^2}
\left[
- \case{1}{4}
{(1+\rho^2)}^2 
dt^2 
+ d\rho^2 + \rho^2 d\theta^2
\right]
\ \ .
\label{sausage-metric}
\end{equation}
We define the time orientation on $\AdSthree$ so that the Killing
vector $\partial_t$ points to the future, and a spatial orientation so
that, for $\rho\ne0$, the pair $(\partial_\rho, \partial_\theta)$ is
right-handed. 

Dropping from (\ref{sausage-metric}) the conformal factor
$4{(1-\rho^2)}^{-2}$ yields a spacetime that can be regularly extended
to $\rho=1$. The timelike hypersurface $\rho=1$ in this conformal
spacetime is by definition the conformal boundary of $\AdSthree$: we
denote this conformal boundary by~$B$. $B$~is a timelike two-torus,
coordinatized by $(t,\theta)$ with the identifications
\begin{equation}
(t,\theta) \sim 
(t,\theta+2\pi)
\sim 
(t+2\pi,\theta)
\ \ , 
\label{boundary-idents}
\end{equation}
and the metric on $B$ is flat, 
\begin{equation}
ds^2 = -dt^2 + d\theta^2
\ \ .
\label{boundary-metric}
\end{equation}
$B$ inherits from $\AdSthree$ a time orientation in which the vector
$\partial_t$ points to the future, and a spatial orientation in which
$\partial_\theta$ points to the right. 

The above definition of a metric on $B$ relies on a particular
coordinate system. The isometries of $\AdSthree$ act on the metric
(\ref{boundary-metric}) as an $\Otwotwo$ group of 
conformal isometries, and the metric on $B$ is thus invariantly
defined only up to such transformations. We therefore understand the
metric (\ref{boundary-metric}) as a representative of its $\Otwotwo$
equivalence class.

The above constructions adapt in an obvious way to the universal
covering space of $\AdSthree$, which we denote by $\CAdSthree$, and to
its conformal boundary, which we denote by~$B_C$. 
When the last identifications in (\ref{sausage-idents}) and
(\ref{boundary-idents}) are dropped, 
the coordinates $(t,\rho,\theta)$
can be regarded as global on $\CAdSthree$, and the coordinates
$(t,\theta)$ can be regarded as global on~$B_C$. $B_C$~has topology
$S^1 \times \BbbR$, and the metric (\ref{boundary-metric}) on $B_C$ is
globally hyperbolic. $B_C$~is time-oriented, with $\partial_t$
pointing to the future, and space-oriented, with $\partial_\theta$
pointing to the right.  The isometries of $\CAdSthree$ clearly induce
conformal isometries of~$B_C$, and the metric on $B_C$ is invariantly
defined only up to these conformal isometries.

\subsection{Spinless nonextremal BTZ hole}
\label{subsec:BTZ}

We now describe the spinless nonextremal BTZ black hole and its
conformal boundary. 

We denote by $\xi_{\rm int}$ and $\eta_{\rm int}$ the Killing vectors
on $\CAdSthree$ that are respectively induced by the Killing vectors
\begin{mathletters}
\begin{eqnarray}
&&
\xi_{\rm emb} := 
- T^1 \partial_{X^1} 
- X^1 \partial_{T^1}
\ \ ,
\\
&&
\eta_{\rm emb} := 
T^2 \partial_{X^2} 
+ X^2 \partial_{T^2}
\ \ ,
\\
\end{eqnarray}
\end{mathletters}
of~$\BbbR^{2,2}$. The conformal Killing vectors that $\xi_{\rm int}$
and $\eta_{\rm int}$ induce on $B_C$ are respectively
\begin{mathletters}
\begin{eqnarray}
&&
\xi := 
\cos t \sin\theta \, \partial_\theta
+ \sin t \cos\theta \, \partial_t
\ \ ,
\\
&&
\eta := 
\cos t \sin\theta \, \partial_t
+ \sin t \cos\theta \, \partial_\theta
\ \ . 
\end{eqnarray}
\end{mathletters}
$\xi_{\rm int}$ and $\eta_{\rm int}$ are clearly mutually orthogonal,
and $\xi$ and $\eta$ are similarly mutually orthogonal.

We denote by $\Dint$ the largest connected region of $\CAdSthree$ that
contains the hypersurface $t=0$ and in which $\xi_{\rm int}$ is
spacelike. As $(\xi_{\rm emb},\xi_{\rm emb}) = {(T^1)}^2 - 
{(X^1)}^2$, we see from (\ref{sausage-coords}) that 
$\Dint$ is isometric to the subset $T^1 > |X^1|$ of the
surface (\ref{embedding-surface}) in~$\BbbR^{2,2}$. 
$\Dint$~intersects every constant $t$ hypersurface for $-\casehalf\pi
< t < \casehalf\pi$, but the only one of these hypersurfaces that is
entirely contained in $\Dint$ is $t=0$. The conformal extension of
$\Dint$ to $B_C$ intersects $B_C$ in the two disconnected diamonds
\begin{mathletters}
\label{DRLdefs}
\begin{eqnarray}
&&
D_R := \left\{ (t,\theta) 
\mid 
\hbox{$0<\theta<\pi$, $|t| < \pi/2 - |\theta - \pi/2|$}
\right\}
\ \ ,
\\
&&
D_L := \left\{ (t, \theta) 
\mid 
\hbox{$-\pi<\theta<0$, $|t| < \pi/2 - |\theta + \pi/2|$}
\right\}
\ \ .
\end{eqnarray}
\end{mathletters}
By construction, $\xi$~is spacelike with respect to the metric
(\ref{boundary-metric}) in $D_R$ and~$D_L$. In the orientations on 
$D_R$ and $D_L$
induced by that on~$B_C$, $\xi$~points to the right in 
$D_R$ and to the left in~$D_L$. $\eta$~is future timelike in~$D_R$,
and past timelike in~$D_L$.

Now, let $a$ be a prescribed positive parameter, and let
$\Gamma_{\rm int}\simeq \BbbZ$ be the group of isometries of
$\CAdSthree$ generated by $\exp(a\xi_{\rm int})$. $\Gamma_{\rm int}$
preserves~$\Dint$, and its action on $\Dint$ is free and properly
discontinuous. The quotient space $\Dint/\Gamma_{\rm int}$ is the
spinless, nonextremal BTZ black hole. The horizon-generating Killing
vector, induced by~$\eta_{\rm int}$, is respectively future and past
timelike in the two exterior regions, and spacelike in the black and
white hole interiors.  The horizon circumference is~$a$, and the mass
is $M=a^2/(32\pi^2 G_3)$, where $G_3$ is the (2+1)-dimensional
Newton's constant.  A~conformal diagram is shown in Figure~1.

\bigskip
\centerline{ \epsfbox{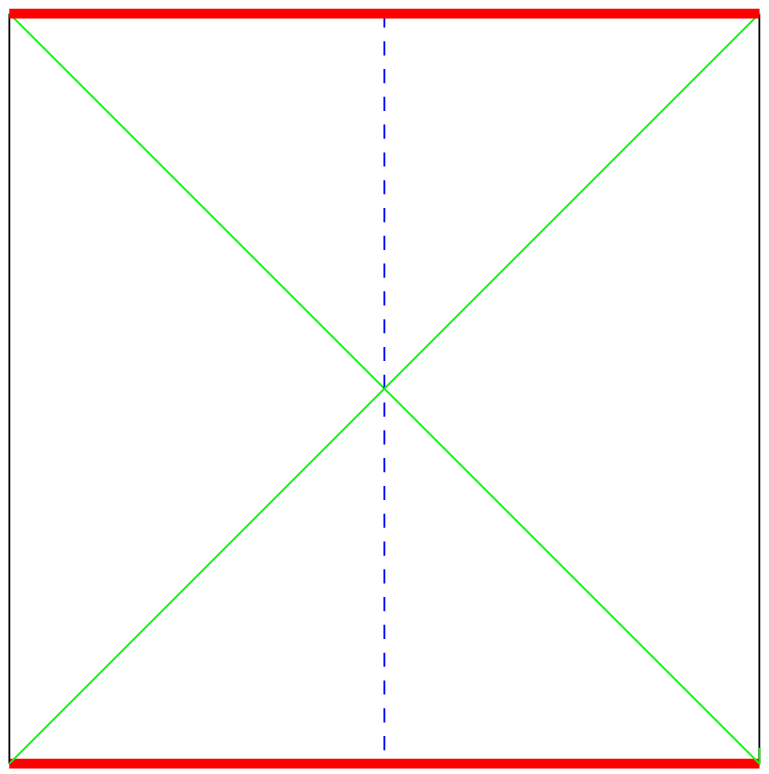}}

\bigskip

{
\narrower\narrower\noindent 
Fig.~1.  
A~conformal diagram of the BTZ
  hole. Each point in the diagram represents a suppressed~$S^1$. The
  involution ${\tilde J}_{\rm int}$ introduced in section
  \ref{subsec:geon} consists of a left-right reflection about the dashed
  vertical line, followed by a rotation
  by $\pi$ on the suppressed~$S^1$.

}

\bigskip

In each of the two exterior regions of the hole, the geometry is
asymptotic to the asymptotic region of $\CAdSthree$. One can therefore
attach to each of the two exterior regions a conformal boundary that is
isometric to~$B_C$. What is important for us is that these conformal
boundaries can be identified as the quotient spaces 
$D_R/\Gamma_R$ and~$D_L/\Gamma_L$, where $\Gamma_R$ and $\Gamma_L$ are
the restrictions to respectively $D_R$ and $D_L$ of the conformal
isometry group of $B_C$ generated by $\exp(a\xi)$
\cite{MAST,horo-marolf-strc}. To see this explicitly, consider~$D_R$,
and cover $D_R$ by the coordinates $(\alpha,\beta)$ defined by
\begin{mathletters}
\label{alphabeta-R}
\begin{eqnarray}
\alpha 
&=&
- \ln \tan\left[(\theta-t)/2\right]
\ \ ,
\\
\beta
&=&
\ln \tan\left[(\theta+t)/2\right]
\ \ .
\end{eqnarray}
\end{mathletters}
Both $\alpha$ and $\beta$ take all real values, and the metric
(\ref{boundary-metric}) on $D_R$ reads
\begin{equation}
ds^2 = - 
\frac{d\alpha d\beta}{\cosh\alpha \, \cosh\beta}
\ \ .
\label{DR-metric1}
\end{equation}
$\partial_\alpha$~and $\partial_\beta$ are future-pointing
null vectors, and
\begin{mathletters}
\begin{eqnarray}
&&
\xi = -\partial_\alpha + \partial_\beta
\ \ ,
\\
&&
\label{eta}
\eta = \partial_\alpha + \partial_\beta
\ \ .
\end{eqnarray}
\end{mathletters}
The generator $\exp(a\xi)$ of $\Gamma_R$ acts in these coordinates as
$(\alpha,\beta)\mapsto(\alpha-a,\beta+a)$, and the metric
(\ref{DR-metric1}) is not invariant under~$\Gamma_R$, but the
conformally equivalent metric 
\begin{equation}
ds^2 = - {\left(\frac{2\pi}{a}\right)}^2 
\, 
d\alpha d\beta
\label{DR-metric1-conf}
\end{equation}
is. The quotient space $D_R/\Gamma_R$, with the metric induced
by~(\ref{DR-metric1-conf}), is thus isometric to $B_C$ with the
metric~(\ref{boundary-metric}). Note that the vector on $D_R/\Gamma_R$
induced by $\eta$ is a future timelike Killing vector in the metric
induced from~(\ref{DR-metric1-conf}), and the isometry with
(\ref{boundary-metric}) takes this vector to the vector
$(2\pi/a)\partial_t$ on~$B_C$.  Similar observations apply to
$D_L/\Gamma_L$, the main difference being that the timelike Killing
vector induced by $\eta$ is now past-pointing, and mapped to
$-(2\pi/a)\partial_t$ under the isometry with~$B_C$.

\subsection{$\RPtwo$ geon}
\label{subsec:geon}

We now turn to the $\RPtwo$ geon. 

Consider on $\CAdSthree$ the isometry $J_{\rm int}$ that is the
composition of $\exp(a\xi_{\rm int}/2)$ and the map
$(t,\rho,\theta)\mapsto(t,\rho,-\theta)$. The group generated by
$J_{\rm int}$ acts on $\Dint$ freely and properly discontinuously. We
define the $\RPtwo$ geon as the quotient space of $\Dint$ under this
group.

As $J_{\rm int}^2=\exp(a\xi_{\rm int})$, $J_{\rm int}$ induces on the
BTZ hole an involutive isometry, which we denote by~${\tilde J}_{\rm
  int}$, and the $\RPtwo$ geon is precisely the quotient space of the
BTZ hole under the $\BbbZ_2$ isometry group generated by~${\tilde
  J}_{\rm int}$. The action of ${\tilde J}_{\rm int}$ on the BTZ hole
is easily understood in the conformal diagram, as shown in Figure 1
 and described in the caption. The conformal diagram
of the $\RPtwo$ geon is shown in Figure~2. It is
clear that the $\RPtwo$ geon is a black hole spacetime with a single
exterior region that is isometric to one exterior region of the BTZ
hole. The geon is time orientable and admits a global foliation with
spacelike hypersurfaces of topology $\RPtwo\setminus${}$\{$point at
infinity$\}$, whence its name; it is, however, not space orientable.
It shares all the local isometries of the BTZ hole. However, as
$J_{\rm int}$ inverts the sign of the Killing vector $\eta_{\rm int}$
on~$\CAdSthree$, $\eta_{\rm int}$ does not induce a globally-defined
Killing vector on the geon, while it does induce a globally-defined
Killing vector on the BTZ hole. 
The quotient construction from the BTZ
hole to the $\RPtwo$ geon is highly similar to the quotient
construction from the Kruskal manifold to the $\RPthree$ geon in four
spacetime dimensions
\cite{Nico,topocen,chamb-gibb,louko-marolf-geon,boersma}.

\bigskip
\centerline{ \epsfbox{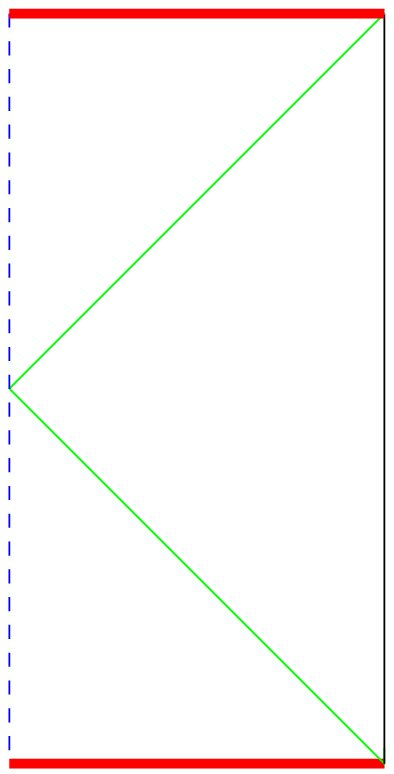}}

\bigskip

{
\narrower\narrower\noindent 
Fig.~2.  
A~conformal diagram of the $\RPtwo$ geon. 
The region not on the dashed
line is identical to that in the diagram of Figure~1,
each point 
representing a suppressed $S^1$ in the spacetime. On the dashed
line, each point in the diagram represents again an $S^1$ in the
spacetime, but with only half of the circumference 
of the $S^1$'s in the
diagram of Figure~1. 

}

\bigskip

The map ${\tilde J}_{\rm int}$ can clearly be extended to the
conformal boundary of the BTZ hole, where it defines an involution
${\tilde J}$ that interchanges the two boundary components. The
conformal boundary of the geon can thus be understood as the quotient
of the conformal boundary of the BTZ hole under the $\BbbZ_2$
generated by~${\tilde J}$. The conformal boundary of the geon is
clearly isometric to a single component of the conformal boundary of
the BTZ hole, and the quotient construction induces on it a time
orientation. Although the boundary of the geon is space orientable,
the quotient construction does not induce a choice for the space
orientation: the reason is that ${\tilde J}$ interchanges the
orientations of the two boundary components of the BTZ hole.

\subsection{Internal dimensions}
\label{subsec:internal-dims}

In string theory, there is interest in spacetimes that are metric
products of $\CAdSthree$ and a compact ``internal'' space. The
quotient constructions of the BTZ hole and the $\RPtwo$ geon can
clearly be extended to such a spacetime and its boundary by taking the
identification group $\Gamma$ 
to act trivially on the internal dimensions.
However, if the internal space admits suitable isometries, other
extensions of interest exist. 

Specifically, if the internal space admits an involutive
isometry~$\Jcompact$, the composition 
$J'$ of $\Jcompact$ and $J_{\rm int}$
is an isometry on the full spacetime, and taking 
$J'$ to generate
the identification group yields a generalization of the $\RPtwo$ geon. An
equivalent construction is to consider first the product spacetime of
the BTZ hole and the internal space, and quotient this by the
$\BbbZ_2$ generated by the 
map $\tilde J'$ that 
is the composition of ${\tilde J}_{\rm int}$
and~$\Jcompact$. The resulting geon has again a single exterior
region. Defining the conformal boundary 
in terms of a conformal
rescaling of the (2+1)-dimensional part of the metric, one finds that
the conformal boundary is connected, and its geometry is that of one
component of the conformal boundary of the BTZ hole with the constant
internal space. Note that while the geometry of the geon depends
on~$\Jcompact$, the geometry of its conformal boundary does not.

In particular, if the internal space is orientable and 
$\Jcompact$ reverses the orientation, the above construction yields a
space-orientable geon. 

The case that will concern us below is a ten-dimensional black hole
spacetime (considered in the context of string theory in Ref.\ 
\cite{MAST}) in which the internal space is the metric product of
$S^3$ and~$T^4$, with a round metric on the former and a flat metric
on the latter. If the $T^4$ factor further factorizes into a metric
product of $S^1$ and~$T^3$, the reflection\footnote{In terms of an
  angular coordinate $\chi$ on the~$S^1$, $\chi\mapsto -\chi$.}  of
the $S^1$ is a nonorientable involutive isometry of the internal
space.  Other nonorientable involutive isometries can be found by
composing this inversion with orientable involutive isometries acting
on the other factors, such as the antipodal map on the~$S^3$. As
smooth orientable quotients of AdS${}_3 \times S^3 \times T^4$, all of
these spacetimes provide exact classical solutions of string theory.

\section{Boundary Conformal Field Theory}
\label{quantum}

We now turn to the question of what sort of quantum
state in the boundary CFT of Maldacena's conjecture \cite{MAL}
is induced by the quotient constructions of section \ref{classic}.
We specialize to the internal space $S^3 \times T^4$ and, 
in order to arrive at an orientable spacetime in which
we might discuss orientable string theory, 
we further assume the metric on the $T^4$ to
factorize in such a way that 
a reflection of an $S^1$ provides an internal
involutive isometry $\Jcompact$ as discussed in
section~\ref{subsec:internal-dims}.
We will not consider the full CFT
suggested by the conjecture but, instead, 
we consider a simplified 
linear field theory which we expect to capture the central
features of interest.

\subsection{Model}
\label{Model}

We consider a set of free scalar fields on the boundary cylinder
$B_C \simeq S^1 \times \R$ of $\CAdSthree$, but with certain
refinements.  The point is that, as discussed in
section~\ref{subsec:internal-dims}, the internal isometry
$\Jcompact$ does not affect how the 
identifications of the full spacetime
$\CAdSthree\times S^3 \times T^4$ project to 
identifications 
of~$B_C$. Nevertheless, 
$\Jcompact$ is expected to affect the full conformal field theory of
Ref.\ \cite{MAL} on~$B_C$.  Thus, our model must contain enough
additional structure to 
faithfully represent the action of~$J_c$.

Recall that the 
$\RPtwo$ geon is a quotient of the BTZ black hole by
the involution $\tilde{J}_{\rm int}$.
Let us take a moment to consider first how this $\Z_2$ quotient 
would be reflected in a boundary CFT\null.  
For a linear field, it is
natural to think of the field 
on both the BTZ hole and geon boundaries as being the {\em same\/} 
operator-valued distribution on the boundary of the BTZ hole, but
merely smeared against different classes of test functions. 
The detailed correspondence
is given by lifting a test function from the geon boundary to 
the BTZ black hole boundary 
and dividing by $\sqrt{2}$ to ensure canonical 
normalization of the field.  Since the geon fields are
embedded in this way in the algebra of BTZ fields, any state on the BTZ
boundary directly induces a state on the geon boundary.
As in
\cite{louko-marolf-geon}, it is sufficient to think of a free
scalar field $\phi_g(x)$ on the geon
boundary as being a symmetrization of the corresponding field 
$\phi_{\rm BTZ}(x)$
residing on the boundary of the BTZ black hole:  
\begin{equation}
\label{sym}
\phi_g(x) = {1 \over {\sqrt{2}}}
\sum_{y \in \rho^{-1} (x)} \phi_{\rm{BTZ}}(y) 
\ \ ,
\end{equation}
where $\rho$ is the covering 
map from the BTZ black hole boundary to 
the geon boundary.

Now, in our construction of the orientable geon from
(BTZ hole)$\times S^3 \times T^4$, the involutive spacetime isometry 
acts on the BTZ hole dimensions by $\tilde{J}_{\rm int}$ and on the
internal toroidal dimensions by reflecting one of the 
$S^1$'s of the~$T^4$. 
Thus, we must include in our model some feature 
that corresponds to 
this internal topology.
We recall that the topology of the internal 
torus is captured \cite{MAL,MAST} by
the fact that the boundary CFT should be a nonlinear
sigma model whose target space is a symmetric product of 
copies of the~$T^4$. A~given $S^1$ factor of the internal
space
is represented as a symmetric product of $S^1$ factors in the target
space of the sigma model.  It is clear that, when acting on the
boundary field theory, the involution that exchanges a point $x$ with
its image $\tilde J x$ should act nontrivially on this part of the
sigma model, reflecting the appropriate $S^1$ factors in the target
space.  This is in direct analogy with the constructions of
\cite{KASI}, where the involutions acted 
only on the internal~$S^3$. 
We will model this feature by replacing the part of the sigma model
associated with the appropriate $S^1$ factors 
by a single scalar field~$\psi$.  
To tighten the analogy with the sigma model, 
one might like to
think of the field space of $\psi$ as compactified to a circle of the
same size as the internal $S^1$ (say, length $2 \pi R$).  However,
this would present
certain problems for our quotient construction.  Setting
aside the reflection of this $S^1$ for the moment, consider the
analogue of (\ref{sym}) for a field $\phi_{\rm BTZ}$ which is periodic
in field space with this period.  The resulting geon field $\phi_g$
would then have period ${2 \pi R}/{\sqrt 2}$.  On the other
hand, the quotient by $\tilde J_{\rm int}$ does not change the size of
the internal $S^1$ factors.  Note that the ${1/\sqrt{2}}$
normalization factor in (\ref{sym}) is fixed by the commutation
relations and the behavior of the Green's functions.

In order to have a faithful representation of the involution $\tilde
J_{\rm int}$, we choose to ignore the compactness of the $S^1$ and to
allow our field $\psi$ to take values on a real line. In analogy with
the $S^1$ it replaces, this line will be reflected through the origin
by the action of the involution on our field theory.  For contrast,
the rest of the sigma model will be replaced by a single free scalar
field $\phi$ whose field space is not affected by the involution.

Our simple model allows exact calculations to be done and 
captures many features of interest.  A notable exception, however, is
that the central charge of our model is~$2$, while that of the CFT
in Maldacena's conjecture is $6Q_1 Q_5$.
This we will correct by hand 
when considering the energies of our states 
in section~\ref{energies}. 

In finding the geon state in the CFT we will proceed as indicated
above, first calculating the boundary state of the BTZ hole in
section~\ref{qBTZ}, and then performing a final identification to
yield the state for the geon in section~\ref{geon}.  
Though the state
on the black hole boundary has been considered in \cite{MAST}, setting
up the BTZ calculation in a different way will make our geon
calculation particularly straightforward.  In addition, we will be
able to see certain effects of the compact boundary that were
neglected in \cite{MAST}.

Sections \ref{qBTZ} and \ref{geon} consider only the oscillator modes
of our scalar fields.  The zero modes are 
more subtle and are treated separately in section~\ref{zero}.  Section
\ref{energies} discusses the energy of 
our states and compares the
result with the mass of the corresponding 
black hole or geon.

\subsection{BTZ black hole state}
\label{qBTZ}
\label{osc}

As discussed in section~\ref{classic}, the BTZ black hole is the
quotient of the region $\Dint \subset \CAdSthree$ under the discrete
isometry group generated by~$\exp(a \xi_{\rm int})$. Similarly, the
boundary of the BTZ hole may be thought of as the quotient of the
region $D_R \cup D_L$ in the boundary $B_C$ of $\CAdSthree$ under the
group generated by~$\exp(a \xi)$. 
We would now like to consider the vacuum state $|0\rangle$ that is
defined on $B_C$ with respect to the timelike Killing
vector~$\partial_t$, and construct the state that $|0\rangle$ induces
on the black hole boundary.
Since the quotient of
$\Dint$ to the BTZ black hole does not act on the internal factors,
the construction is identical for both $\phi$ and $\psi$ and the
discussion below applies to either field.

Recall that the null coordinates $\alpha$ and $\beta$
(\ref{alphabeta-R}) define a conformal mapping of $D_R$ onto Minkowski
space (and similarly for~$D_L$). In terms of this Minkowski space, the
map $\exp(a \xi)$ is just a spatial translation, and when the overall
scale of the metric is chosen as in~(\ref{DR-metric1-conf}), the
proper distance of the translation is~$2\pi$. Thus, these
identifications enact the usual compactification of Minkowski space to
$S^1 \times \R$.  The effect of the compactification on the scalar
field theory is merely to remove all modes that are not appropriately
periodic and to reinterpret the periodic modes, which are not
normalizable on $D_R$ (or~$D_L$), as normalizable modes on the
cylinder.

The nontrivial part in this construction is that, as noted in
section~\ref{subsec:BTZ}, the timelike Killing vectors on the two
boundary components of the BTZ black hole do not lift to the timelike
Killing vector $\partial_t$ on~$B_C$: the future timelike Killing
vector on the boundary component arising from $D_R$ lifts to
$[a/(2\pi)]\eta$, and that on the boundary component arising from
$D_L$ lifts to $-[a/(2\pi)]\eta$. Thus, in order to interpret 
the state induced by $|0\rangle$ on
the BTZ hole boundary
in terms of the BTZ particle modes, 
we must first write the state induced by
$|0\rangle$ on $D_R\cup D_L$ in terms of continuum-normalized particle
states that are positive frequency with respect to $\eta$ on $D_R$ and
with respect to $-\eta$ on~$D_L$.  This calculation is quite similar
to expressing the Minkowski vacuum in terms of Rindler particle modes
(see e.g.\ \cite{takagi-magnum,birrell-davies,wald-qft}).

To begin, consider the mode functions 
\begin{mathletters}
\label{rindler-modes}
\begin{eqnarray}
&&
u^{R}_{\omega, \epsilon}
:= 
\frac{1}{\sqrt{4\pi\omega}} 
{\left[
\tan\biglb((\theta - \epsilon t)/2 \bigrb)
\right]}^{i\epsilon\omega}
\ \ ,
\\
&&
u^{L}_{\omega, \epsilon}
:= 
\frac{1}{\sqrt{4\pi\omega}} 
{\left[
\tan\biglb((-\theta + \epsilon t)/2 \bigrb)
\right]}^{-i\epsilon\omega}
\ \ ,
\end{eqnarray}
\end{mathletters}
where $\omega>0$, the index $\epsilon$ takes the values~$\pm1$, and
the modes with superscript $R$ ($L$) have support in $D_R$~($D_L$).
The $R$-modes are eigenfunctions of the vector field $\eta$ on $D_R$
with eigenvalue~$\omega$, and the $L$-modes are similarly
eigenfunctions of the vector field $-\eta$ on $D_L$ with
eigenvalue~$\omega$. The modes are continuum orthonormal on
$D_R \cup D_L$. 
The modes with $\epsilon=1$ are right-moving
and those with $\epsilon=-1$ are left-moving, in the orientations on 
$D_R$ and $D_L$
induced by that on~$B_C$.\footnote{As mentioned after
  equations~(\ref{DRLdefs}), in this orientation 
$\xi$ is right-pointing in $D_R$ and left-pointing in~$D_L$. } 
These properties for
the $R$-modes become explicit by writing the modes in terms of the
null coordinates $(\alpha,\beta)$ (\ref{alphabeta-R}) on $D_R$ as
$u^{R}_{\omega, +} = {(4\pi\omega)}^{-1/2} e^{-i\omega\alpha}$ and
$u^{R}_{\omega, -} = {(4\pi\omega)}^{-1/2} e^{-i\omega\beta}$.
Analogous expressions hold for the $L$-modes on~$D_L$. 

Let now $|0 \rangle_{\rm osc}$ stand for the vacuum of the non-zero
modes induced on $D_R
\cup D_L$ by~$|0 \rangle$, and let $|0 \rangle_u$ stand for the vacuum
of the $u$-modes~(\ref{rindler-modes}). We need to express
$|0\rangle_{\rm osc}$ in terms of $|0 \rangle_u$ and the excitations
associated with the $u$-modes. To this end, we follow the method of
Unruh \cite{unruh-magnum} and build from the $u$-modes and their
complex conjugates a complete set of linear combinations, called
$W$-modes, that are bounded analytic functions in the lower half
complex $t$ plane. By construction, the $W$-modes are purely positive
frequency with respect to~$\partial_t$, and they thus share the
vacuum~$|0 \rangle_{\rm osc}$. The relevant Bogoliubov transformation
can then be simply read from expressions of the $W$-modes. 

Analytically continuing the $u$-modes (\ref{rindler-modes}) between
$D_R$ and $D_L$ in the lower half of the complex $t$ plane, we find
that a complete set of $W$-modes is 
\begin{mathletters}
\label{W-modes}
\begin{eqnarray}
&&
W^{(1)}_{\omega,\epsilon}
= \frac{1}{\sqrt{2\sinh(\pi\omega)}}
\left(
e^{\pi\omega/2} 
u^{R}_{\omega, \epsilon}
+ 
e^{-\pi\omega/2} 
\overline{u^{L}_{\omega, \epsilon}}
\right)
\ \ ,
\\
&&
W^{(2)}_{\omega,\epsilon}
= \frac{1}{\sqrt{2\sinh(\pi\omega)}}
\left(
e^{\pi\omega/2} 
u^{L}_{\omega, \epsilon}
+ 
e^{-\pi\omega/2} 
\overline{u^{R}_{\omega, \epsilon}}
\right)
\ \ ,
\end{eqnarray}
\end{mathletters}
where $\omega > 0$ and $\epsilon = \pm 1$. The creation and
annihilation operators $a_{\omega,\epsilon}^{\dagger (i)}$,
$a_{\omega,\epsilon}^{(i)}$ for the $W$-modes are thus related to the
creation and annihilation operators $b_{\omega,\epsilon}^{\dagger
{L,R}}$, $b_{\omega,\epsilon}^{L,R}$ for the $u$-modes by
\begin{mathletters}
\label{a-vs-b}
\begin{eqnarray}
&&
a^{(1)}_{\omega,\epsilon}
= \frac{1}{\sqrt{2\sinh(\pi\omega)}}
\left(
e^{\pi\omega/2} 
b^{R}_{\omega, \epsilon}
- 
e^{-\pi\omega/2} 
b^{\dagger L}_{\omega, \epsilon}
\right)
\ \ ,
\\
&&
a^{(2)}_{\omega,\epsilon}
= \frac{1}{\sqrt{2\sinh(\pi\omega)}}
\left(
e^{\pi\omega/2} 
b^{L}_{\omega, \epsilon}
- 
e^{-\pi\omega/2} 
b^{\dagger R}_{\omega, \epsilon}
\right)
\ \ .
\end{eqnarray}
\end{mathletters}
To relate the vacua, one notices that equations (\ref{a-vs-b}) can
be written as \cite{israel-vacuum}
\begin{mathletters}
\begin{eqnarray}
a^{(1)}_{\omega, \epsilon} 
&=& \exp(-iK) b^R_{\omega,\epsilon} \exp(iK)
\ \ ,
\\
a^{(2)}_{\omega, \epsilon} 
&=& 
\exp(-iK) b^L_{\omega,\epsilon} \exp(iK)
\ \ ,
\end{eqnarray}
\end{mathletters}
where $K$ is the (formally) Hermitian operator
\begin{equation}
K = i 
\sum_{\epsilon}
\int_0^\infty d \omega  
\, r_\omega
\left(
b^{\dagger R}_{\omega, \epsilon}
b^{\dagger L}_{\omega, \epsilon} -
b^{R}_{\omega, \epsilon}
b^{L}_{\omega, \epsilon} \right)
\end{equation}
and $r_\omega$ is defined by 
\begin{equation}
\tanh(r_\omega) = \exp(-\pi \omega)
\ \ .
\end{equation}
The vacuum $|0 \rangle_{\rm osc}$, 
annihilated by the $a^{(i)}_{\omega, \epsilon}$, 
is therefore related to the vacuum $|0 \rangle_u$, 
annihilated by the $b_{\omega,\epsilon}^{L,R}$, through 
\begin{equation}
| 0 \rangle_{\rm osc} = \exp(-iK) |0 \rangle_{u}
\ \ .
\end{equation}
In terms of the normalized $q$-particle states 
$|q\rangle_{\omega,\epsilon}^{R}$ 
($|q\rangle_{\omega,\epsilon}^{L}$) associated with the modes
$u_{\omega,\epsilon}^{R}$ 
($u_{\omega,\epsilon}^{L}$), this relation reads
\begin{equation}
|0\rangle_{\rm osc}
=
\prod_{\omega > 0, \epsilon}
\left(
{1 \over \cosh(r_\omega)}
\sum_{q=0}^\infty
\exp(-\pi\omega q)
|q\rangle_{\omega,\epsilon}^{R} 
|q\rangle_{\omega,\epsilon}^{L}
\right)
\ \ .
\label{vacads-expanded}
\end{equation}

We can now pass from the field theory on $D_R \cup D_L$ to the field
theory on the boundary of the BTZ hole. Let us refer to the non-zero
modes of the field as oscillator modes. For the oscillator modes, the
effect of the periodic identifications is simply to replace the
continuous index $\omega$ by the discrete values $\omega = \omega_n :=
2 \pi n/a$, where the index $n$ takes values in the positive integers,
and to change the normalization factor 
in the $u$-modes (\ref{rindler-modes}) from 
${(4\pi\omega)}^{-1/2}$ to ${(4\pi n)}^{-1/2}$. 
For the oscillator modes, we then obtain
from $|0\rangle_{\rm osc}$ (\ref{vacads-expanded}) the state
\begin{equation}
|{\rm BTZ}\rangle_{\rm{osc}}
=
\prod_{n > 0, \epsilon}
\left(
{1 \over \cosh(r_{\omega_n})}
\sum_{q=0}^\infty
\exp(-\pi\omega_n q)
|q\rangle_{n,\epsilon}^{R} 
|q\rangle_{n,\epsilon}^{L}
\right)
\ \ .
\label{vacbtzhole-expanded}
\end{equation}
$|{\rm BTZ}\rangle_{\rm{osc}}$ clearly lies in the Fock space 
${\cal H} = {\cal H}_{R} \otimes {\cal H}_{L}$, 
where ${\cal H}_{R}$ 
and ${\cal H}_{L}$ are the Hilbert spaces of the oscillator
modes of the scalar field on the respective $S^1 \times \R$ conformal
boundary components of the BTZ hole.  Note that $|{\rm
  BTZ}\rangle_{\rm{osc}}$ (\ref{vacbtzhole-expanded}) is properly
normalized and that it may be written as $|{\rm{BTZ}} \rangle_{
  \rm{osc}} = \exp (-iK_{\rm{BTZ}}) |0 \rangle_{ \rm{osc}}$, where
\begin{equation}
\label{JBTZ}
K_{\rm{BTZ}} = i \sum_{n>0,\epsilon} r_{\omega_n}
\left(
b^{\dagger R}_{n, \epsilon}
b^{\dagger L}_{n, \epsilon} -
b^{R}_{n, \epsilon}
b^{L}_{n, \epsilon} \right)
\ \ . 
\end{equation}

When both fields $\phi$ and $\psi$ are considered (together with the
zero mode states discussed below in section~\ref{zero}), $|\rm{BTZ}
\rangle_{\rm osc}$ gives the BTZ black hole quantum state in our model
theory on $(S^1 \times \R) \cup (S^1 \times \R)$.  We see from
(\ref{vacbtzhole-expanded}) that $|\rm{BTZ} \rangle_{\rm{osc}}$
contains pairwise correlations between modes residing on the two
boundary components, and when the modes on one component are traced
over, the resulting state on the other component is thermal.
The expectation value of any operator associated with only one
boundary component is thus identical to the expectation value in a
thermal state. This is in particular true of the stress-energy tensor.

Finally, to identify the temperature of the effective thermal state on
a single boundary component, we recall that the discretized $u$-modes
are eigenfunctions of $\pm\eta$ with eigenvalue~$\omega_n$.  The
expression (\ref{vacbtzhole-expanded}) therefore implies that the
thermal state has temperature $1/(2\pi)$ with respect to~$\pm\eta$,
which translates into the temperature 
$a/4\pi^2$ with respect to the
Killing vector $\partial_t$ in the form (\ref{boundary-metric}) of the
boundary metric. This is the usual Hawking temperature of (the
interior of) the BTZ black hole with respect to a Killing time
coordinate that agrees with our $t$ on the boundary \cite{BTZ}.
We shall discuss the energy expectation values further
in section \ref{energies} after having first addressed the zero modes.

\subsection{Geon state}
\label{geon}

We now construct the state of the oscillator modes in our 
model on the boundary of the 
$\RPtwo$ geon. The zero modes 
will be discussed below in section~\ref{zero}.

Let $\rho_L$ and $\rho_R$ be the restrictions of the covering map
$\rho$ of the BTZ hole boundary over 
the boundary of the geon to the left and
right components of the BTZ hole boundary. The geon boundary fields
$\psi_{g}$ and $\phi_{g}$ are then 
related to the BTZ fields by 
\begin{mathletters}
\label{geonfields}
\begin{eqnarray}
\psi_{g}(x) &:=& 
\frac{1}{\sqrt{2}}
\left[ \psi_{\rm BTZ}\biglb(\rho_R^{-1}(x)\bigrb) 
- \psi_{\rm BTZ}
\biglb(\rho_L^{-1}(x)\bigrb) 
\right]
\ \ ,
\\
\phi_{g}(x) &:=& 
\frac{1}{\sqrt{2}}
\left[ \phi_{\rm BTZ}\biglb(\rho_R^{-1}(x)\bigrb) 
+ \phi_{\rm BTZ}
\biglb(\rho_L^{-1}(x)\bigrb) 
\right]
\ \ .
\label{geonfields-phi}
\end{eqnarray}
\end{mathletters}
The argument $x$ of $\psi_{g}(x)$ and $\phi_{g}(x)$ 
takes values in a single copy of
$S^1 \times \R$, but the field operators act in the Hilbert space 
${\cal H}_{\rm{BTZ}}$ of the BTZ boundary theory.
As the BTZ black hole state is symmetric with respect to the
sign of~$\psi$, the geon state will not depend on which boundary
component is called left or right.

What we wish to do is to calculate the restriction of the
BTZ state to the algebra generated by the
fields~(\ref{geonfields}).  Note that the restriction of
a pure state to a subalgebra is not necessarily pure.  Thus, a
priori, the result could be either a pure state or a mixed state.

We proceed by introducing two more 
fields, $\tilde \psi_g$ and~$\tilde \phi_g$, through 
\begin{mathletters}
\label{fakefields}
\begin{eqnarray}
\tilde \psi_{g}(x) &:=& 
\frac{1}{\sqrt{2}}
\left[ \psi_{\rm BTZ}\biglb(\rho_R^{-1}(x)\bigrb) 
+ \psi_{\rm BTZ}
\biglb(\rho_L^{-1}(x)\bigrb) 
\right]
\ \ ,
\\
\tilde \phi_{g}(x) &:=& 
\frac{1}{\sqrt{2}}
\left[ \phi_{\rm BTZ}\biglb(\rho_R^{-1}(x)\bigrb) 
- \phi_{\rm BTZ}
\biglb(\rho_L^{-1}(x)\bigrb) 
\right]
\ \ .
\label{fakefields-phi}
\end{eqnarray}
\end{mathletters}
These fields again live on $S^1 \times \R$ and 
act in the Hilbert space 
${\cal H}_{\rm{BTZ}}$. The definitions 
(\ref{geonfields}) and (\ref{fakefields}) 
amount to writing the two fields 
$\bigl\{
\psi_{\rm BTZ}, 
\phi_{\rm BTZ}
\bigr\}$
on the two-component BTZ boundary as the four fields 
$\bigl\{
\psi_{g},
\phi_{g},
\tilde \psi_{g},
\tilde \phi_{g}
\bigr\}$
on a single copy of $S^1 \times \R$.  
The Hilbert space 
${\cal H}_{\rm{BTZ}}$ then factors as
${\cal H}_{\rm {BTZ}} = {\cal H}_g \otimes \tilde {\cal H}_{g}$,
where
${\cal H}_g$ is the Hilbert space of 
the geon fields (\ref{geonfields})
while 
$\tilde {\cal H}_g$ is the Hilbert 
space of the fields~(\ref{fakefields}).
The desired state in the geon boundary theory (${\cal H}_g$)
then follows by tracing over the
Hilbert space $\tilde {\cal H}_{g}$.   
In fact, taking this trace will
be trivial as we will see that the state 
$|\rm{BTZ} \rangle_{\rm{osc}}$ 
is a tensor product state, containing
no correlations between 
${\cal H}_{g}$ and~$\tilde {\cal H}_{g}$.
That this must be so follows from the observation that 
$|{\rm BTZ }\rangle_{\rm osc}$ 
contains only two-particle correlations. Since this
state vector is invariant 
under the operation of interchanging the right and left boundary 
components, it can only contain correlations between fields of the same
parity under this operation. 

Note that the only difference between the fields $\phi$ and $\psi$ is
in the signs in (\ref{geonfields}) and~(\ref{fakefields}), and that
interchanging the tilded geon boundary
fields for the untilded ones is
equivalent to interchanging $\phi$ for $\psi$ on the BTZ
boundary. 
Thus, the state of $\psi_g$ is identical
to the state of $\tilde \phi_g$ (on the BTZ boundary) and the state of
$\tilde \psi_g$ is 
identical to 
that of $\phi_g$.  As a result, it will again be
sufficient to 
treat only one of the fields 
$\phi$ and $\psi$
explicitly.
We choose the field~$\phi$, 
and then read off the state of $\psi$ from the results.

Consider thus the field~$\phi$. 
A~complete orthonormal basis of positive frequency oscillator
modes on the geon boundary is given by the functions 
$U_{n, \epsilon}(x) := u^{R}_{n, \epsilon} \biglb( \rho_R^{-1}(x) \bigrb)$,
which are the pushforward to the geon of the modes (\ref{rindler-modes}) 
on the right BTZ boundary.
We denote the annihilation and creation operators associated
with the field $\phi_g$ in this basis by $d_{\phi,n,\epsilon}$
and~$d^\dagger_{\phi,n,\epsilon}$, and those associated with the field
$\tilde \phi_g$ by $d_{\tilde \phi,n,\epsilon}$ and~$d^\dagger_{\tilde
  \phi,n,\epsilon}$. As the properties of the involution 
$\tilde{J}$ (whose quotient of the BTZ boundary yields the geon
boundary) imply 
that the pullback of these modes to the left BTZ boundary differs
from (\ref{rindler-modes}) by a $\pi$ rotation (and a definition of left-
and right-moving), 
we have
$u^{L}_{n, \epsilon}
\biglb(
\rho_L^{-1}(x)
\bigrb)
= 
{(-1)}^n
U_{n, -\epsilon}(x)$, and
we then find from (\ref{geonfields-phi}) and
(\ref{fakefields-phi}) the relations 
\begin{mathletters}
\label{d-vs-b}
\begin{eqnarray}
d_{\phi,n,\epsilon} 
&=& 
{1 \over
 \sqrt{2} } \left[ b^R_{\phi, n, \epsilon} 
+ (-1)^n b^L_{\phi, 
n, -\epsilon} \right]
\ \ ,
\\
d_{\tilde \phi,n,\epsilon} 
&=& 
{1 \over
 \sqrt{2} } \left[ b^R_{\phi, n, \epsilon} 
- (-1)^n b^L_{
\phi, n, -\epsilon} \right] 
\ \ . 
\end{eqnarray}
\end{mathletters}
Using~(\ref{d-vs-b}), 
the operator $K_{\rm{BTZ}}$ (\ref{JBTZ}) can be
written in the form 
\begin{equation}
\label{diag}
K_{\rm{BTZ}} = i \sum_{n = 1}^{\infty} (-1)^n  
\left[
\left( d^\dagger_{\phi,n,+} d^\dagger_{\phi,n,-} - 
d_{\phi,n,+} d_{\phi,n,-} \right)
- \left (d^\dagger_{\tilde \phi,n,+} 
d^\dagger_{\tilde \phi,n,-} - 
d_{\tilde\phi,n,+} d_{\tilde \phi,n,-} \right) 
\right]
\ \ .
\end{equation}
The parts of (\ref{diag}) referring to $\phi_g$ and $\tilde\phi_g$
each have the same form as the terms in~(\ref{JBTZ}),  
apart from some changes of signs.
By the same
methods as in section~\ref{qBTZ}, we can therefore write $|{\rm
  BTZ}\rangle_{\rm{osc}}$ in terms of the 
normalized $q$-particle
states $|q \rangle_{\phi,n,\epsilon}$ and $|q \rangle_{\tilde
  \phi,n,\epsilon}$ associated with the operators
$d^\dagger_{\phi,n,\epsilon}$ and~$d^\dagger_{\tilde \phi,n,\epsilon}$
as
\begin{equation}
\label{diagstate}
|{\rm BTZ}\rangle_{\rm{osc}}
= 
\prod_{n > 0 , \sigma \in \{ \phi, \tilde \phi \} }
\left(
{1 \over \cosh(r_{\omega_n})}
\sum_{q=0}^\infty (-1)^{nq} (-1)^{q [s(\sigma)]}
\exp(-\pi\omega_n q)
|q\rangle_{\sigma, n,+}
|q\rangle_{\sigma, n,-}
\right)
\ \ ,
\end{equation}
where we have defined $s(\phi) := 0$ and $ s(\tilde \phi) :=1$. This
means in particular that $|{\rm BTZ} \rangle_{\rm{osc}}$ is a direct
product of a state in ${\cal H}_g$ with a state in~$\tilde {\cal
  H}_g$.  The restriction of (\ref{diagstate}) to the
field $\phi_g$ therefore yields the geon oscillator state for~$\phi$.

For the field~$\psi$, the calculation is similar except in that the
tilded and untilded fields are interchanged. The geon oscillator
state for $\psi$ can therefore be read off from the restriction of
(\ref{diagstate}) to~$\tilde \phi_g$. Thus, defining 
$s(\psi) :=1$, the geon
oscillator state including both fields is
\begin{equation}
|{\rm geon} \rangle_{\rm {osc}} 
= \prod_{n >0, \sigma \in \{\psi, \phi \} }
\left(
{1 \over \cosh (r_{\omega_n})}
\sum_{q=0}^\infty
\exp(-\pi\omega_n q)
(-1)^{nq} (-1)^{q[s(\sigma)]}
|q\rangle_{\sigma, n,+}
|q\rangle_{\sigma, n,-}
\right)
\ \ , 
\label{vacbtzhole-proj}
\end{equation}
which is a normalized pure state in~${\cal H}_g$.

The correlations between the right-movers and the
left-movers exhibited in (\ref{vacbtzhole-proj}) 
are similar to the
correlations found in scalar field theory on 
the (interior of the) $\RPthree$ geon spacetime 
and on an analogous Rindler-type spacetime
in Ref.\ \cite{louko-marolf-geon}.\footnote{In 
Ref.\ \cite{louko-marolf-geon}, the counterparts of the 
minus signs appearing in 
(\ref{vacbtzhole-proj}) were encoded in the phase choices for
the mode functions.} 
We shall discuss this phenomenon further in section~\ref{disc}.

\subsection{Zero modes}
\label{zero}

In our calculations of the zero mode states below, we replace
the oscillator modes on the CAdS${}_3$, BTZ hole, and geon boundaries
by modes of finite frequency~$\Omega$.  We then take the limit $\Omega
\rightarrow 0$ to give the state for the actual zero modes of our massless
fields.  Now, the reader may be concerned by the fact that modifying the
zero modes on the boundary of the BTZ hole 
affects not only the zero
mode on~$B_C$; it will modify the oscillator modes as well.  However, 
the procedure below may be thought of as a condensed version of the
more manifestly self-consistent procedure of giving our fields a finite
mass $m$ (which of course affects all of the modes together) and then
taking the $m \rightarrow 0$ limit.  In this longer and more complicated
calculation, one would compute the state of the massive field on $D_L
\cup D_R$ in terms of modes that are positive frequency with respect to
$\eta$ on $D_R$ and $-\eta$ on $D_L$ and then take the $m \rightarrow 0$
limit to yield the state of the massless field (zero mode and all) on
$D_L \cup D_R$.  The massless field state can then be compactified
as before.  It will be clear that the calculation below gives identical
results.

Now, the fact that the
zero mode energy eigenstates of 
a free field on $S^1 \times \R$ 
are not normalizable will lead to some subtleties in our argument.  
In particular, 
the ground state $|0\rangle$ from which the BTZ and geon states
are induced is non-normalizable.  Thus, the
BTZ and geon states are unlikely to be normalizable, and a limit
taken in the Hilbert space topology will not be useful.  We will
proceed by considering the states as tempered distributions on the
zero mode configuration space.  Note that, in the topology of tempered
distributions, a suitably rescaled version of the harmonic oscillator
ground state does in fact converge to the free particle ground state
$\delta  (p)$, where $p$ is the free particle momentum.  
The rescaling is necessary since $\delta (p)$ is not a 
normalizable state in the Hilbert space.  Our limit will require
a rescaling of the state as well, and, 
for this reason, we induce the BTZ and
geon states from the state $|0 \rangle_\Omega$, which is ${1/
{\sqrt{\pi \Omega}}}$ times the normalizable ground state for the 
frequency $\Omega$ zero modes.  More will be said about the precise form
of this rescaling 
at the end of the calculation.  
Below, we first take the limit in the sense of (smooth) functions on
the configuration space.  We then note that this convergence is
sufficiently uniform to guarantee that the same limit is given by the
topology of tempered distributions.  

Let us begin by replacing the zero mode on the boundary of 
$\CAdSthree$ by the 
$\Omega > 0$ oscillator mode $(4 \pi \Omega)^{-1/2} e^{-i \Omega t}$, 
and similarly on the two components of the BTZ boundary and on the geon
boundary.  In this case, the BTZ zero modes are associated
with the modes
\begin{mathletters}
\begin{eqnarray}
u^R_{\Omega} &=& {1 \over {\sqrt{4 \pi \Omega}}} 
\, e^{-i\Omega (\alpha + \beta)/2} 
= 
{1 \over {\sqrt{4 \pi \Omega}}} \left[
{{\tan \biglb( (\theta - t)/2 \bigrb)} \over
{\tan \biglb( (\theta + t)/2 \bigrb)} }  \right]^{i\Omega/2}
\ \ , 
\\
u^L_{\Omega}  
&=&
{1 \over {\sqrt{4 \pi \Omega}}} \left[
{{\tan \biglb( (-\theta - t)/2 \bigrb)} \over
{\tan \biglb( (-\theta + t)/2 \bigrb)} } \right]^{i\Omega/2}
\ \ ,
\end{eqnarray}
\end{mathletters}
in the domains ${D}_{L,R}$ on the boundary of $\CAdSthree$.  We refer
to the creation and annihilation operators for such modes as
$b^{\dagger,L,R}_{\Omega}$ and $b^{L,R}_{\Omega}$.  As before, we need
only explicitly calculate the BTZ state for one of our scalar fields.

The calculation proceeds much as in section \ref{qBTZ}, except
that the zero mode does not have separate left- and right-moving
parts.  Thus, the operator $K_{\rm{BTZ},\Omega}$ which relates the
zero mode part of
the regulated BTZ zero mode state $|\rm{BTZ} \rangle_\Omega$ 
to that of the vacuum $| 0 \rangle_\Omega$  on the BTZ boundary
takes the form
\begin{equation}
\label{KOm}
K_{\rm{BTZ},\Omega} = i r_\Omega \left( d^{\dagger L}_{\Omega} 
d^{\dagger R}_{\Omega}  
- d^L_{\Omega} d^R_{\Omega}
\right)
\end{equation}
with the corresponding form 
\begin{equation}
\label{BOm}
|{\rm BTZ }\rangle_{\Omega} = \left( { {1} \over {\sqrt{\pi \Omega}}} \right)
{1 \over {\cosh(r_\Omega)}} \sum_{q=0}^\infty
\exp(-\pi \Omega q) |q   \rangle^R_\Omega |q \rangle^L_\Omega
\end{equation}
for the zero mode state in terms of normalized
$q$-particle states.  Here, the fact that our state $|0 \rangle_\Omega$
is $( \pi \Omega)^{-1/2}$ times a normalized state can
be seen explicitly.

We must now take the limit as the frequency $\Omega$
is sent to zero.  
To proceed, recall that the 
states $|q \rangle^{L,R}_\Omega$ may be thought of as the normalized
occupation number states for a
harmonic oscillator on the real line.  We therefore introduce
the usual position states $\{ |x \rangle ^{L,R} \}$ [normalized to 
${}^{L,R}\langle
x | x' \rangle^{L,R} = \delta(x-x')$] and momentum states
$\{ | p \rangle^{L,R} \}$ 
[normalized to ${}^{L,R}\langle p | p' \rangle^{L,R} 
 = \delta(p-p')$]
for this particle, as well as the tensor
products $|x_L,x_R \rangle = |x_L \rangle_L \otimes |x_R \rangle_R$ and
$|p_L,p_R \rangle = |p_L \rangle \otimes |p_R \rangle$.
In the limit $\Omega \rightarrow 0$, the occupation number states
must go over to energy eigenstates of the free particle.  Moreover, since
states with $q = 2k$ have positive parity, they
must be proportional to the positive
parity states $ | p \rangle^{L,R}_+ \equiv
(|p \rangle^{L,R} + |-p \rangle^{L,R}) 
/\sqrt{2}$ for the appropriate momentum $p$
in the limit of small $\Omega$. 
Similarly, odd states with $q = 2k+1$ must become proportional
to $ | p \rangle^{L,R}_- \equiv
(|p \rangle^{L,R} - |-p \rangle^{L,R}) /\sqrt{2}$. 

To fix this remaining constant of proportionality, 
consider the even wave functions \cite{Weissbluth}
\begin{equation}
{}^{L,R}\langle x | 2k \rangle^{L,R}_\Omega 
= \left({{ \Omega} \over {\pi}} \right)^{1/4}  
[2^{2k} (2k)!]^{-1/2} H_{2k} 
(x \sqrt{\Omega }) e^{-x^2 \Omega/2}
\end{equation}
of the oscillator states.  
Here $H_n$ is the Hermite polynomial of order~$n$.
Of course, since the states $|2k \rangle_\Omega^{L,R}$ are normalized, 
the wave function at any point $x$ vanishes as $\Omega \rightarrow 0$.
In contrast, we have $\langle x | p \rangle_+ = (1/\sqrt{\pi}) \cos(px)$.
Thus, if we fix a compact set $\Delta \subset \R$, in the limit of 
small $\Omega$ with $E = (2k + {1 \over 2}) \Omega$ held fixed
we have 
\begin{equation}
\label{wfr}
{}^{L,R}\langle x | 2k \rangle^{L,R}_\Omega 
\rightarrow \left({{\Omega} \over { \pi}} 
\right)^{1/4}  [2^{2k} 
(2k)!]^{-1/2} H_{2k} (0) \sqrt{\pi} \left(  {}^{L,R}\langle x | p 
\rangle^{L,R}_+ \right)
\end{equation}
uniformly on $\Delta$.  Using the fact that, in the $\Omega 
\rightarrow 0$ limit at fixed~$x$, the coefficient in (\ref{wfr}) 
is independent of $k$ and 
the creation operator goes
over to $-i (2 \Omega)^{-1/2}$ times the momentum operator, one can
show that the relative normalizations of the wave functions
${}^{L,R}\langle x|2k+1 \rangle^{L,R}_\Omega$ and ${}^{L,R}\langle
x | p \rangle^{L,R}_-$ are the same up to a factor of $-i$:
\begin{equation}
\label{odd}
{}^{L,R}\langle x | 2k +1 \rangle^{L,R}_\Omega 
\rightarrow -i \left({{\Omega} \over { \pi}} 
\right)^{1/4}  [2^{2k} 
(2k)!]^{-1/2} H_{2k} (0) \sqrt{\pi} \left(  {}^{L,R}\langle x | p 
\rangle^{L,R}_- \right) .
\end{equation}

Let us now evaluate the part of the wave function 
$\langle x_L, x_R | {\rm BTZ} 
\rangle_{ \Omega}$
that comes from harmonic oscillator states with energies $E_{q}$ in a
small interval $E - \delta E/2 < E_{q} < E + \delta E/2$ in the limit of
small $\Omega$.  We include both even parity ($q=2k$) and odd parity
($q=2k+1$) states.  The associated momentum interval is $\delta p
= p^{-1} \delta E$ 
where $p = \sqrt{2E} = \sqrt{4k \Omega }$.
  For fixed $x_{L,R} \ll 1/\sqrt{\delta E}$ the values of the
wave functions ${}^{L,R}\langle x_{L,R}|2k \rangle^{L,R}_\Omega$ 
for the allowed values of $k$
are nearly identical and
are given by (\ref{wfr}). 
Now, the even Hermite polynomials at zero are given 
\cite{arfken} by $H_{2k} (0) = (-1)^k {{(2k)!}/{k!}}$.
Using Stirling's approximation 
$n! = (\sqrt{2 \pi n}) n^n e^{-n}$, and the
fact that there are $\delta E/2\Omega$ 
states of each parity in the allowed energy range, 
the contribution of these
states is
\begin{equation}
 \sqrt 2 \exp(-\pi p^2/2) \left(
\langle x_L,x_R |p,-p \rangle +
\langle x_L,x_R |p,-p \rangle \right) (\delta p).   
\end{equation}
Thus, summing over all such intervals $\delta p$ and considering
all $x_L,x_R \in \R$ gives the zero frequency state of the zero
mode for either $\phi$ or $\psi$:
\begin{equation}
\label{phiz}
|{\rm BTZ} \rangle_{0} = \sqrt 2 \int_{-\infty}^\infty 
dp \ \exp(-\pi p^2/2)
|p \rangle_L |-p \rangle_R \ \ .
\end{equation}
Although we have taken this limit in the topology
of pointwise convergence on $\R^2$, the exponential
cutoff $\exp(-\pi \Omega q)$ and exponential falloff of the oscillator
wave functions in the momentum representation can be 
used to show that that (\ref{phiz})
is in fact the limit of $|{\rm BTZ} \rangle_{\Omega}$ in the sense
of tempered distributions.

Thus, expression (\ref{phiz}) is the zero mode
state on the BTZ boundary.  We see that the trace over either
boundary component yields a thermal state at the same temperature
as for the oscillators.  As expected, neither (\ref{phiz}) nor the state
traced over one component is normalizable.
We see that this is due
to the precise correlation of the momenta on the right and left in
(\ref{phiz}).  It turns out that something similar must happen
whenever the zero mode spectrum is continuous.  This is because
the state $|0 \rangle$
from which the BTZ state is induced is invariant under the action
of the Killing field $\eta$ (\ref{eta}).  On the BTZ boundary
this corresponds to the action of the difference $H_R- H_L$ between
the right and left boundary Hamiltonians.  Thus, $H_R-H_L$
will annihilate the BTZ state.   But, when the zero mode spectrum is
continuous, $H_R-H_L$ has no normalizable eigenstates.

To arrive at the geon boundary state for, say, $\phi$, 
we need only introduce
the basis $|p,\tilde p \rangle \rangle =
| {{(p + \tilde p)}/ {\sqrt{2}}}, 
 {{(p - \tilde p)} / {\sqrt{2}}}\rangle$ in terms of eigenvalues 
$p, \tilde p$ of the momenta conjugate to the zero modes of $\phi_g$ and 
$\tilde \phi_g$.  The BTZ state (\ref{phiz}) for $\phi$ may be written
\begin{equation}
\label{ppt}
|{\rm BTZ} \rangle_{\phi,0} = \sqrt 2 \int
dp d \tilde p \ \delta(p) \exp(-\pi \tilde p^2/4)
|p , \tilde p \rangle  \rangle \ \ .
\end{equation}
As in section~\ref{geon}, there are no correlations between
$\tilde \phi_g$ and~$\phi_g$.
Thus, we may read off from (\ref{ppt}) the geon boundary states
for both $\psi$ and~$\phi$. 
The (normalized) geon state for $\psi$ is given simply by the factor
$2^{-1/4} \exp(-\pi \tilde p^2/4)$ corresponding to 
$\tilde \phi_g$ in~(\ref{ppt}), 
while the geon state for $\phi$
is given by the other factor $ 2^{3/4} \delta(p)$.  
Note that the zero mode of $\phi$ is in its ground state.
One might expect this result to be maintained if
the field space of $\phi$ could be compactified, in which
case the ground state would of course be normalizable. 

At this point, a comment is in order on the form of the factor $(\pi 
\Omega)^{-1/2}$ by which we needed to rescale the normalized
ground state.  The reader will note that the distribution $\delta(p)$
is in fact the limit
(as a distribution over the configuration space )
of $\left ( {4 / {\Omega \pi}} \right)^{1/4}$
times the normalized Harmonic oscillator ground state.  
Thus, the limit
of $|0 \rangle_\Omega$ 
as $\Omega \rightarrow 0$ is not~$|0 \rangle$, but is
larger by $(4 \pi \Omega)^{1/4}$.  That this
extra rescaling is necessary results from the fact that the
fluctuations of $p_L + p_R$ in our state are much smaller than the
fluctuations of the momentum in the ground state of the harmonic
oscillator.

The geon state for $\psi$ is normalizable but the corresponding state
for $\phi$ is not.  We note that the $\psi$ zero mode state can
in fact be calculated without dealing with 
distributions at all.  To do so, 
one first writes the operator 
$K_{\rm{BTZ},\Omega}$ (\ref{KOm}) in terms of creation
and annihilation operators for the 
zero modes of $\psi_g$ and
$\tilde \psi_g$.  As usual, this gives a sum of two operators, one
involving $\psi$ and one involving~$\tilde \psi$.  Letting the 
exponential of ($-i$ times) the 
$\psi$ part act on a normalized
vacuum state gives a one-parameter
family of normalized states
that converges to the above result 
{\em in the Hilbert space norm\/} as
$\Omega \rightarrow 0$.
We consider this an important check on our use of distributions above.
Note that, since $H_L-H_R = {1 \over 2} (P_L^2 - P_R^2)$ must annihilate
the full state 
$|{\rm BTZ} \rangle_{\psi, 0}$, 
it then follows that $\tilde
\psi_g$ (and therefore $\phi_g)$ 
is in the zero momentum state.

\subsection{Energy expectation values}
\label{energies}

We now examine the expectation value of the energy (expected energy)
in our quantum states.  
For the BTZ black hole, we consider the Hamiltonian associated with
a single boundary component (say, the one on the right).  Note
that it is the ADM Hamiltonian of a single asymptotic region
that gives the classical mass of the black hole \cite{BTZ}.
Note also that the black hole mass is associated with the Killing
field $\partial_t$ of the boundary metric (\ref{boundary-metric}).
Thus, we consider the notion of energy defined by this vector field.
The thermal behavior noted
in section \ref{osc} is therefore associated with a temperature 
$T = {a / {4 \pi^2}}$.

Because the total energy is a sum of the Hamiltonians for the left-
and right-moving modes separately, it is apparent from an 
examination of (\ref{vacbtzhole-expanded}) and (\ref{vacbtzhole-proj}) 
that the expected 
energy of the oscillator modes is {\em exactly} the same for
the BTZ and geon states.  The extra minus signs in (\ref{vacbtzhole-proj}) 
do not affect the expectation value and the Hamiltonian of, say, 
the right-moving modes on our boundary component
does not care whether a right-moving state there
is correlated with a mode on 
another boundary component or with a left-moving
mode on the same boundary component.  
In both cases, the result is
just the expected energy in a thermal state 
of temperature $a/ 4 \pi^2$.
The same is also true of the 
stress-energy tensor.

For a zero mode on a component of the BTZ boundary, one sees
from the regulated expression (\ref{BOm}) that the state again acts like a
thermal state at the same temperature. While such a 
state is not normalizable in
the $\Omega \rightarrow 0$ limit, 
the expectation value of the energy associated with $\partial_t$
is finite and equal to ${a /{4\pi^2}}$.
On the geon boundary, the expected energy of the $\psi$ zero mode is
${a /{2\pi^2}}$ while that of the $\phi$ zero mode vanishes.  

For the oscillator modes of a free scalar field on a cylinder, the
energy expectation value in a thermal state is well known
\cite{birrell-davies}. The circumference of our cylinder
(\ref{boundary-metric}) is~$2\pi$, and the temperature 
is~${a /{4\pi^2}}$: 
with these parameters, one finds from the general formulas given in
Ref.\ \cite{birrell-davies} that the energy of our oscillator state
relative to the ground state (Casimir) energy is
\begin{equation}
\label{therm}
{1 \over 2 } \sum_{m=1}^\infty 
\frac{1}{\sinh^2 (2 \pi^2 m /a)}
\ \ .
\end{equation}

To connect 
these results within our model with the full CFT, we 
recall that our model theory has central charge $2$ while the
full CFT of \cite{MAL} has central charge $6Q_1Q_2$.  
Since the field space of $\psi$ 
represents one of the four $S^1$ factors of the internal torus 
(and thus one fourth of the non-linear sigma model), 
we may expect that a central charge of ${3 \over 2}Q_1Q_5$ is associated
with the part of the sigma model that is similar to~$\psi$, while
the remaining ${9 \over 2} Q_1 Q_5$ is associated with fields 
similar to~$\phi$.  We therefore model the energy of the full
theory with ${3 \over 2} Q_1Q_5$ copies of $\psi$ and 
${9 \over 2} Q_1 Q_5$ 
copies of~$\phi$.  

As noted above, the energy is the same for both $\phi$ and $\psi$ on
the BTZ boundary.  Thus, the total energy there is given by $6Q_1Q_5$ times
the expression (\ref{therm}) plus ${{3Q_1Q_5 a}/ {2 \pi^2}}$ for
the zero modes.  For our $\RPtwo$ geon, the zero mode of $\phi$
is in its ground state but the zero mode of $\psi$ has energy
${a /{2\pi^2}}$.  Thus, the total energy is $6Q_1 Q_5$ times
expression (\ref{therm}) plus ${{3aQ_1Q_5}/ {4 \pi^2}}$.  In the limit
$a \gg 1$ (large black hole and large temperature of the thermal states)
the zero mode correction is negligible and (using $\sum_{m=1}^\infty m^{-2}
= \pi^2/6$) the expected energy reduces
to $ {{a^2 Q_1 Q_5} /{8\pi^2}} $.  In our notation (and for the
spinless case we consider), 
$\tilde T_+$ and $\tilde T_-$ of \cite{MAST} are given by
$\tilde T_+ = \tilde T_- = {a /{4 \pi^2}}$.
We also note
that all energies in \cite{MAST} were computed with respect to a Killing
vector field 
that corresponds to $R^{-1} \partial_t$, where $R$ is defined
in \cite{MAST}.   With this
understanding, our result in this limit agrees with \cite{MAST} (which
did not take into account the discrete nature of the field modes).

Note that we have not set the three-dimensional
Newton's constant $G_3$ equal
to one and, in fact, it is fixed \cite{MAL,MAST}
by the relation between
the Anti-de Sitter space and the central charge of the CFT\null.  
Since we have set the length scale $l$ of Anti-de Sitter space to one, 
$G_3^{-1}$ is $4Q_1Q_5$ (times the string scale).  
It follows that the energy of our CFT state also
agrees with the classical black hole mass  
${{a^2} /{32 \pi^2 G_3}}$
for $a \gg 1$.
This observation was made in \cite{MAST} (in which this
limit was taken implicitly) in the context of BTZ holes. Note that 
taking $a \gg 1$ gives the 
limit in which the black hole is much larger
than the radius of curvature of the AdS space, and it 
is in this regime that the energy of a thermal
bath\footnote{Or, an `approximately thermal' pure state such as
the $\RPtwo$ geon analogue of the 
scalar field vacuum on the $\RPthree$
geon constructed 
in \cite{louko-marolf-geon}.} 
required to maintain equilibrium
would be small compared to the mass of the black hole.

\section{Swedish geon}
\label{Swedish}

In this section we investigate a CFT on the boundary of another
geon-type, single-exterior, (2+1)-dimensional black hole spacetime:
the spinless black hole with spatial topology
$T^2\setminus${}$\{$point at infinity$\}$ constructed in Ref.\ 
\cite{ABBHP} and analyzed in detail in Ref.\ \cite{ABH}. We refer to
this spacetime as the Swedish geon.\footnote{References 
\cite{ABBHP,ABH,horo-marolf-strc} used the 
term ``wormhole.''}  We will not be able to obtain the CFT
state as explicitly as for the $\RPtwo$ geon, but we can reduce the
problem of finding this state to a mathematical problem involving
certain automorphic functions. We will also be able to
contrast the correlations present in the Swedish geon state to those
present in the $\RPtwo$ geon state.

As the Swedish geon is space and time orientable, we consider as a
model theory a single conformal scalar field $\phi$ that lives on the
boundary of the spacetime.

Let us briefly recall the construction of the Swedish geon and its
conformal boundary \cite{ABBHP,ABH}. Let 
$\xi_{\rm int}$ and ${\tilde\xi}_{\rm int}$ 
be on $\CAdSthree$ the Killing vectors
respectively induced by the Killing vectors
\begin{mathletters}
\begin{eqnarray}
&&
\xi_{\rm emb} := 
- T^1 \partial_{X^1} 
- X^1 \partial_{T^1}
\ \ ,
\\
&&
{\tilde\xi}_{\rm emb} := 
- T^1 \partial_{X^2} 
- X^2 \partial_{T^1}
\ \ ,
\end{eqnarray}
\end{mathletters}
of~$\BbbR^{2,2}$. The conformal Killing vectors induced on $B_C$ are
respectively 
\begin{mathletters}
\begin{eqnarray}
&&
\xi := 
\cos t \sin\theta \, \partial_\theta
+ \sin t \cos\theta \, \partial_t
\ \ ,
\\
&&
{\tilde\xi} := 
- \cos t \cos\theta \, \partial_\theta
+ \sin t \sin\theta \, \partial_t
\ \ . 
\end{eqnarray}
\end{mathletters}
Note that~$\xi_{\rm int}$ and $\xi$ are as in
section~\ref{classic}. The Swedish geon is now defined as the quotient
of a certain subset of $\CAdSthree$ under the infinite 
discrete group generated
by $A_{\rm int}:= \exp(-a\xi_{\rm int})$
and 
$B_{\rm int}:= \exp(-a{\tilde\xi}_{\rm int})$, 
where the parameter $a$
satisfies $\sinh(a/2)>1$. The geon is space and time orientable, 
it admits a global foliation with spacelike hypersurfaces of
topology 
$T^2\setminus${}$\{$point at infinity$\}$, 
and it has a single exterior
region isometric to that of a spinless nonextremal BTZ black hole with
horizon circumference $\gamma := 4 \arccosh[\sinh^2(a/2)]$. The
singularities, and the exotic topology, are hidden behind the
horizon. 

It follows from the above that the conformal boundary of the Swedish
geon consists of just one copy of the conformal boundary of
$\CAdSthree$. As explained in detail in Ref.\ \cite{ABH}, this
boundary emerges from the boundary $B_C$ of the original $\CAdSthree$
as the quotient of a set $D\subset B_C$ under the discrete group
$\Gamma^S$ generated by $A:= \exp(-a\xi)$ and $B :=
\exp(-a{\tilde\xi})$.  $D$~consists of a countable number of
disconnected diamonds, each of them the domain of dependence of an
open interval in the $t=0$ circle: the end points of the intervals are
at the fixed points of $\Gamma^S$ on this circle.

Recall from section \ref{classic} that it was possible to describe 
the $\RPtwo$
geon boundary by considering just one of the diamonds of $D_R\cup
D_L$, and taking its quotient under the identification subgroup 
that maps this diamond to itself. A~similar description is
possible for the Swedish geon boundary \cite{ABH}. Among the 
countably many
diamonds
constituting~$D$, let $D_1$ be the one that intersects $t=0$ in the
interval $(\pi/4)-\arccos[C/(\sqrt{2}S)] < \theta < (\pi/4) +
\arccos[C/(\sqrt{2}S)]$, where 
$S := \sinh(a/2)$ and
$C := \cosh(a/2)$. It can be shown that the only elements of
$\Gamma^S$ that leave $D_1$ invariant are powers of $G_1 :=
ABA^{-1}B^{-1}$, and that the boundary of the Swedish geon is the
quotient of $D_1$ under the $\BbbZ$ generated by~$G_1$.

Now, $G_1$ can be written as $G_1 = \exp(\gamma\xi_1)$, where 
\begin{equation}
\xi_1 := 
\frac{1}{\sqrt{S^2 - 1}}
\left[ 
C \partial_\phi + S \left( {\tilde\xi} - \xi \right)
\right] 
\ \ .
\end{equation}
$\xi_1$ is a conformal Killing vector on~$B_C$, and its fixed points
at $t=0$ are precisely at the corners of~$D_1$, at $\theta = (\pi/4)
\pm \arccos[C/(\sqrt{2}S)]$. The conformal Killing vector $\xi_1$ is
thus analogous to the conformal Killing vector $\xi$ on $D_R$ in
section~\ref{classic}.  In particular, $D_1$ 
admits a future timelike
conformal Killing vector orthogonal to~$\xi_1$, analogous to $\eta$
on~$D_R$, and this conformal Killing vector defines the positive and
negative frequencies on the Swedish geon boundary.

Consider our conformal scalar field $\phi$ on~$D_1/G_1$. We 
introduce on $D_1$ the null coordinates $(u,v)$: 
\begin{mathletters}
\begin{eqnarray}
&&u := t - [\theta - (\pi/4)]
\ \ ,
\\
&&v := t + [\theta - (\pi/4)]
\ \ ,
\end{eqnarray}
\end{mathletters}
which cover $D_1$ with $|u| < \arccos[C/(\sqrt{2}S)]$ and $|v| <
\arccos[C/(\sqrt{2}S)]$.  In analogy with~(\ref{rindler-modes}), one
finds that a complete orthonormal basis for the oscillator modes of
$\phi$ on~$D_1/G_1$, positive frequency with respect to the geon
boundary time, is
\begin{mathletters}
\label{Sgeon-modes}
\begin{eqnarray}
&&
U_{n,+} := 
\frac{1}{\sqrt{4\pi n}} 
\left[
\frac{V + \tan(u/2)}{V-\tan(u/2)}
\right]^{-2\pi i n/\gamma}
\ \ ,
\label{Sgeon-modes-u}
\\
&&
U_{n,-} := 
\frac{1}{\sqrt{4\pi n}} 
\left[
\frac{V+\tan(v/2)}{V-\tan(v/2)}
\right]^{-2\pi i n/\gamma}
\ \ ,
\end{eqnarray}
\end{mathletters}
where $n$ takes values in the positive integers and 
\begin{equation}
V := \sqrt{
\frac{\sqrt{2}S - C}{\sqrt{2}S + C}}
\ \ .
\end{equation}
The subscript $+$ ($-$) yields the right-moving (left-moving) modes. 
When the metric on the 
geon boundary is written as
in~(\ref{boundary-metric}), the frequency with respect to $\partial_t$
is just~$n$. The vacuum of the modes (\ref{Sgeon-modes}) is therefore
the usual vacuum on the geon boundary for the oscillator modes of the
field: we denote this vacuum by~$|0\rangle_U$. 
The usual vacuum for the zero modes is again nonnormalizable;
from now on we 
restrict the discussion to the oscillator modes. 

We would now like to use Unruh's analytic continuation method
\cite{unruh-magnum} to find the oscillator mode state
$|\hbox{S-geon}\rangle_{\rm osc}$ that is induced on the boundary of the geon
by the usual oscillator mode 
vacuum $|0\rangle_{\rm osc}$ on~$B_C$. 
This means that we must form from the $U$-modes
(\ref{Sgeon-modes}) and their complex conjugates linear combinations,
the $W$-modes, that satisfy two requirements. First, when analytically
continued to the lower half of the complex planes in $u$ and~$v$, the
$W$-modes must be bounded analytic functions: this guarantees that
they are purely positive frequency linear combinations of the modes
that define~$|0\rangle_{\rm osc}$. Note that the $W$-modes may have
singularities at certain real values of $u$ and~$v$, but apart from
these singularities, the analytic continuation defines them as
functions on all of $B_C$ and not just in the diamond $D_1\subset
B_C$.

Second, the $W$-modes must accommodate the fact that the geon boundary
field operator on $D$ is constructed by averaging $\phi$ over
$\Gamma^S$, as in~(\ref{geonfields-phi}). This means that the
$W$-modes must be invariant over~$\Gamma^S$, while each of the
$U$-modes (\ref{Sgeon-modes}), when analytically continued from $D_1$
to~$D$, is individually invariant only under the subgroup of
$\Gamma^S$ generated by~$G_1$.

It is easy to see that $\Gamma^S$ takes $u$-independent functions into
$u$-independent functions and similarly $v$-independent functions into
$v$-independent functions. The $W$-modes can therefore be divided into
right-movers, constructed from $\left\{U_{n,+}\right\}$ and their
complex conjugates, and left-movers, constructed from
$\left\{U_{n,-}\right\}$ and their complex conjugates. For
concreteness, consider the right-movers. To put the problem into a
mathematically familiar form, we replace $u$ by the coordinate 
$z := \cot\biglb((u/2) + 3\pi/8\bigrb)$. 
In terms of~$z$, $B_C$~corresponds to the 
compactification of
the real line, and $D_1$ is covered by the interval
$k_-<z<k_+$, where 
$k_\pm := e^{-a/2}\left( S \pm \sqrt{S^2-1} \right)$. 
Analytic continuation of $u$
into the lower half-plane is equivalent to analytic continuation of
$z$ into the upper half-plane. The generators $A$ and $B$ act on $z$
as fractional linear transformations whose matrices [which, as
elements of ${\rm PSL}(2,\BbbR)\simeq{\rm SL}(2,\BbbR)/(\pm\openone)$,
are defined only up to the overall sign] are
\begin{mathletters}
\begin{eqnarray}
&&
{\hat A} = 
\pm 
\left(
\begin{array}{lll}
 e^{-a/2}  && 0 \\  
0 
&& e^{a/2} \\
\end{array}
\right)
\ \ ,
\\
&&
{\hat B} = 
\pm 
\left(
\begin{array}{lll}
 \cosh(a/2)  && \sinh(a/2) \\  
\sinh(a/2)
&& \cosh(a/2) \\
\end{array}
\right)
\ \ .
\end{eqnarray}
\end{mathletters}
The $W$-modes are thus the bounded analytic functions in the upper $z$
half-plane that are invariant under the group generated by ${\hat A}$
and~${\hat B}$.\footnote{Note that ${\hat A}$
and ${\hat B}$ are boosts with magnitude~$a$, the fixed points of
${\hat A}$ are at $z=0$ and $z=\infty$, and the fixed points of ${\hat
  B}$ are at $z=\pm1$. This makes the $W$-modes
  automorphic functions \cite{terras1} on the noncompact Riemann
  surface that is isomorphic to the time-symmetric hypersurface in the
  geon {\em spacetime\/}. For a fundamental domain for this Riemann
  surface, see Ref.\ \cite{ABBHP}.}
Expressing the modes $\left\{U_{n,+}\right\}$ 
(\ref{Sgeon-modes-u}) in terms of~$z$, we see that finding the
Boboliubov transformation reduces to 
finding the coefficients $a_n$ in the
expansions 
\begin{equation}
W = \sum_{n\ne0} \frac{a_n}{\sqrt{4\pi|n|}} 
\, 
\left[
\frac{(1 + k_+)}{\sqrt{2}}
\left( \frac{z - k_-}{-z + k_+} \right) 
\right]
^{2\pi n i/\gamma}
\ \ ,
\label{Sge-Wexp}
\end{equation}
where the terms with positive $n$ come from the $U_{n,+}$
and the terms with negative $n$ come from the complex
conjugates. 
Note that, by construction, each term on the right-hand-side of
(\ref{Sge-Wexp}) is invariant under the fractional linear
transformation 
${\hat A}{\hat B}{\hat A}^{-1}{\hat B}^{-1}$, 
whose fixed points are at $z=k_\pm$. 

We shall not pursue the analysis further here, but we make one
speculative comment. 
On the real axis, both 
${\hat A}$ and ${\hat B}$ map the interval $k_-<z<k_+$ 
completely outside this interval. If 
$k_-<z<k_+$, 
both ${\hat A}$ and ${\hat B}$ thus 
take each term in the sum (\ref{Sge-Wexp}) to a term whose magnitude
differs by the factor $e^{-2\pi^2n/\gamma}$. This suggests (but
certainly does not prove) that if the sum is to be invariant, the
coefficients $a_n$ should 
be exponentially increasing in~$n/\gamma$. 
If this is true, comparison with the relative weights of the terms in
(\ref{W-modes}) suggests (but again certainly does not prove) that
$|\hbox{S-geon}\rangle_{\rm osc}$ might appear in some respects as a
thermal state in a temperature proportional to~$\gamma$.  We leave the
examination of these speculations subject to future work.

\section{Discussion}
\label{disc}

We have seen that, in our model, an $\RPtwo$ geon corresponds to a pure state
of finite energy on the CAdS${}_3$
boundary.  In particular, 
these states contain correlations between the right- and 
left-moving sectors such that, when one of these sectors is traced over, 
the other sector is left in a thermal state.  The expectation value of the
Hamiltonian in an $\RPtwo$ geon state is exactly the same as the expectation
value of the Hamiltonian (for either of the boundary components) 
in the corresponding
BTZ black hole state, and this value agrees with the
the classical mass of the 
spacetimes in the limit where the
black hole is much larger than the length scale of the AdS space.

In our model, the zero mode parts of our states $|{\rm BTZ }\rangle$ and
$|{\rm geon }\rangle$ were not normalizable.
This was due to the noncompact range of our fields and the resulting
continuous spectrum of the zero mode Hamiltonian.  For the same
reason, the ground state $| 0 \rangle$ of our model theory is again
non-normalizable.  Since our final states were in fact induced
from the ground state, it is
no surprise that our construction failed to generate normalizable states.
Indeed, the surprise is that the field 
$\psi$ is in a normalizable state
on the geon boundary.  In all cases, we
arrive at a generalized state that may be expressed in the usual way
in terms of distributions.

While we were not able to complete
a corresponding analysis of the Swedish geon states, the
fact that these black holes have only a single asymptotic region 
(with topology
$S^1 \times \R$) 
leads one to once again expect pure states. 
Still, calculating one of these states would be of interest
as it is far from clear what sort of correlations it would contain.
In particular, in contrast with an $\RPtwo$ geon, 
the identifications that lead
to a Swedish geon act separately on the right- and 
left-moving parts of the
CFT\null.  Thus, there should be no correlations between right- and 
left-moving
modes and the correlations must take a rather different
form than for an $\RPtwo$ geon.  
Nonetheless, the hypothesis that a Swedish geon is `not too far' 
from a thermal state is supported by the behavior of the (as yet
formal) Unruh modes of section \ref{Swedish} under analytic continuation.
Another motivation for studying the Swedish geon is that, since 
the identifications that yield an orientable Swedish geon need not
act on the internal factors, such geons may more readily allow a treatment
of fields with compact target spaces. 

One might also try to generalize our calculation to
higher-dimensional single-exterior locally AdS black holes constructed
from the two-exterior locally AdS black holes of Refs.\ 
\cite{ABHP,BLP,RM,MB} via a suitable involution.  However, such
single-exterior black holes are of somewhat less interest 
as their asymptotic topology is
always different from the asymptotic topology of AdS space.  Thus,
the black hole and the AdS space will 
in any case not correspond to quite the same boundary field
theory.

Of course, the real interest is to extrapolate our results to the more
complicated CFT which forms the basis of Maldacena's duality
conjecture \cite{MAL}.  The main difficulty with our model was that we
were unable to capture the compact nature of the moduli space of this
theory.  It is unclear to us to what extent quotients of the type
described here can actually be carried out in a nonlinear field
theory.

It is natural to assume, however, that the major qualitative difference
between our model and the full theory is that 
the BTZ and geon states become normalizable, 
placing the fields $\phi_g$
and $\tilde \psi_g$ in their ground states.
Certainly, we would once again expect an $\RPtwo$ geon to
be associated with a pure state 
(of finite energy) in the usual Hilbert space.
It should contain correlations of the 
sort found here, between identical right- and left-moving modes, again 
defining a thermal state with respect to either the right- or 
left-moving sector alone. 
Although we have not discussed
fermions in our model, if they were included with antiperiodic
boundary conditions the associated geon state would be interpreted 
as an excited state of the
CFT vacuum representing AdS space.  On the other hand, fermions
may also be included with periodic boundary conditions, in which case
the ground state of the CFT is associated \cite{MAST} with the $M=0$
BTZ black hole and our geon is a corresponding excited state.  
These two cases would correspond to string
theories twisted in different ways around the nontrivial topology
of the black hole throat.

As excited states, the $\RPtwo$ geon states are certainly not
invariant under time translations.  In fact, an inspection of
(\ref{vacbtzhole-proj}) shows that they are not even stationary.  This
is in accordance with the fact that 
the timelike Killing field in the exterior region of the geon
spacetime cannot be extended to a globally-defined Killing field on
all of the spacetime. 

Now, in the theory considered here, all of the modes (except the zero
modes) of our scalar fields are periodic in time with a common period.
Thus, the oscillator part of our geon state is actually periodic in
time.  This is not a feature of the classical geon spacetime and it is
{\em not\/} a feature that one would expect to survive in the full
boundary CFT\null.  Indeed, already in describing the original AdS
space we see that this periodicity must be broken (by the anomalous
dimensions of certain operators in the CFT \cite{LUMA}) if the
boundary theory is to describe aperiodic processes.

In order to see what should be expected when this periodicity is
broken, we note that the construction of the geon boundary state is in
direct parallel with the construction given in
\cite{louko-marolf-geon} 
of vacua on the entire asymptotically flat
$\RPthree$ geon spacetime and on an analogous Rindler-type spacetime.
In those cases, it was found that the correlations between field modes
became unobservable by localized detectors far from the preferred time
$t=0$ and that the state behaved for many purposes as a thermal 
({\it i.e.}, mixed) state.  Clearly, we expect parallel results here.  It
is true that, since any correspondence between the boundary CFT and
the bulk string theory will be nonlocal, the relevance of local
detectors on the boundary is unclear.  However, one still expects that
the geon state will, in an appropriate sense, approximate the BTZ
state over a single boundary component at early and late times.

Perhaps one of the most interesting aspects of our calculation is
the way in which 
the reflection of an internal $S^1$ is represented 
in the boundary quantum state.  Our results are similar to
those of \cite{KASI}, in that the identifications on the internal
dimensions are reflected in certain symmetries of the CFT quantum
states.  In our model, this involved the state of the scalar
field~$\psi$, which was associated with the $S^1$ factor on which our
spacetime identifications act.  By including more of the full
nonlinear sigma model and thus capturing more of the internal
dimensions, we could arrive at similar states that correspond to
other orientable $\RPtwo$-like geons with different internal spaces,
including orbifolds.  In contrast, the complicated topology of the
Swedish geons is associated only with 
the AdS factor of the spacetime 
(and not with the internal compact dimensions).  
It would therefore be interesting to probe this issue further through
a full calculation of a Swedish geon boundary state.

Finally, we note that if $\tilde \psi_g$ were the physical field on
the geon, the oscillator state would differ only by removing the
factors of $(-1)^{s(\psi)}$.  Placing the zero mode in its ground
state, one constructs in this way a state that one is tempted to
associate with string theory on a non-orientable $\RPtwo$ geon.  Thus,
one might speculate that it may be possible to describe states of
non-orientable string theory in terms of the same CFT Hilbert space.
Whether or not this happens in the full theory or is merely an
artifact of our model must, of course, be left for future studies.

\acknowledgements

We would like to thank 
Vijay Balasubramanian, 
Ingemar Bengtsson, 
Steve Carlip, 
Nico Giulini, 
Gary Horowitz, 
Albion Lawrence, 
Juan Maldacena, 
and 
Max Niedermaier 
for discussions and correspondence. 
D.M.~was supported in part by NSF grant PHY-9722362 and
research funds provided by Syracuse University. 
D.M.~would like to thank the
Max-Planck-Institut f\"ur Gravitationsphysik for their hospitality
during the final stages of this work.


\begin{references}

\bibitem{HS} 
G.~T. Horowitz and A.~Strominger, 
Nucl.\ Phys.\ {\bf B360}, 197 (1991). 

\bibitem{Dark} 
G.~T. Horowitz, 
% ``The dark side of string theory,''
in {\it String theory and quantum gravity '92}, 
Trieste 1992, Proceedings, 
55-99 (hep-th/9210119). 

\bibitem{bekenstein1}
J.~D. Bekenstein,
Lett.\ Nuovo Cimento {\bf 4}, 737 (1972).

\bibitem{bekenstein2}
J.~D. Bekenstein,
Phys.\ Rev.\ D {\bf 9}, 3292 (1974).

\bibitem{Hawk} 
S.~W. Hawking, 
Commun.\ Math.\ Phys.\ {\bf 43}, 199 (1975).

\bibitem{SV} 
A.~Strominger and C.~Vafa, 
Phys.\ Lett.\ B {\bf 379}, 99 (1996). 
(hep-th/9601029) 

\bibitem{G}
J.~P. Gauntlett, 
% ``Intersecting Branes'',
% {\it preprint\/}
hep-th/9705011.

\bibitem{GGPT}
J.~P. Gauntlett, 
G.~W. Gibbons, 
G.~Papadopoulos, and 
P.~K. Townsend,
Nucl.\ Phys.\ {\bf B500}, 133 (1997). 
(hep-th/9702202) 

\bibitem{AS}
A.~Sen,
Phys.\ Rev.\ D {\bf 53}, 2874 (1996). 
(hep-th/9511026) 

\bibitem{BKOP}
E.~Bergshoeff, 
R.~Kallosh, 
T.~Ortin, and 
G.~Papadopoulos,
Nucl.\ Phys.\ {\bf B502}, 149 (1997). 
(hep-th/9705040) 

\bibitem{RPS}
M. de~Roo, 
S.~Panda, and 
J.~P. van~der Schaar, 
Phys.\ Lett.\ B {\bf 426}, 50 (1998). 
(hep-th/9711160)

\bibitem{MAL}  
J.~Maldacena,  
% ``The Large N Limit of Superconformal Field Theories 
% and Supergravity'', 
% {\it preprint\/} 
Adv.\ Theor.\ Math.\ Phys.\ {\bf 2}, 231 (1997). 
(hep-th/9711200)

\bibitem{GUKL} 
I.~Klebanov, 
Nucl.\ Phys.\  {\bf B496}, 231 (1997) 
(hep-th/9702076); 
S.~Gubser and I.~Klebanov, 
Phys.\ Lett.\ B {\bf 413}, 41 (1997)
(hep-th/9708005); 
M.~Douglas, 
J.~Polchinski, and 
A.~Strominger,
J.~High Energy Phys.\  {\bf 12}, 3 (1997)
(hep-th/9703031). 

\bibitem{evidence}
S.~Ferrara, 
C.~Fronsdal, and A.~Zaffaroni, 
Nucl.\ Phys.\ {\bf B532}, 153 (1998)
(hep-th/9802203); 
O.~Aharony,
Y.~Oz, 
and 
Z.~Yin, 
Phys.\ Lett.\ B {\bf 430}, 87 (1998)
(hep-th/9803051); 
J.~de Boer, 
hep-th/9806104.

\bibitem{MAST}  
J.~Maldacena and A.~Strominger, 
% ``AdS${}_3$ Black Holes and a 
% Stringy Exclusion Principle'',
% {\it preprint\/} 
hep-th/9804085.

\bibitem{BBG}  
K.~Behrndt, 
I.~Brunner, and 
I.~Gaida, 
% ``Entropy and Conformal Field 
% Theories of AdS${}_3$ Models'', 
% {\it preprint\/} 
Phys.\ Lett.\ B {\bf 432}, 310 (1998). 
(hep-th/9804159)

\bibitem{TL}  
T.~Lee, 
% ``The Entropy of the BTZ Black Hole 
% and AdS/CFT Correspondence'', 
% {\it preprint\/} 
hep-th/9806113.

\bibitem{BDHM} 
T.~Banks, 
M.~R.~Douglas, 
G.~T.~Horowitz, and 
E.~Martinec, 
hep-th/9808016.

\bibitem{BKLT} 
V.~Balasubramanian, 
P.~Kraus, A.~Lawrence, 
and 
S.~P.~Trivedi, 
hep-th/908017.

\bibitem{keskivak} 
E.~Keski-Vakkuri, 
hep-th/9808037.

\bibitem{BTZ} 
M.~Ba\~nados, 
C.~Teitelboim, and J.~Zanelli, 
Phys.\ Rev.\ Lett.\ {\bf 69}, 1849 (1992)
(hep-th/9204099); 
M.~Ba\~nados, M.~Henneaux,
C.~Teitelboim, and J.~Zanelli, 
Phys.\ Rev.\ D {\bf 48}, 1506 (1993)
(gr-qc/9302012). 

\bibitem{ABBHP} 
S.~\AA{}minneborg, 
I.~Bengtsson, 
D.~R. Brill, 
S.~Holst, and
P.~Peld\'an, 
Class.\ Quantum Grav.\ {\bf 15}, 627 (1998). 
(gr-qc/9707036) 

\bibitem{ABH} 
S.~\AA{}minneborg, 
I.~Bengtsson, and  S.~Holst,  
% ``A~Spinning Anti-de Sitter Wormhole'', 
% {\it preprint\/} 
gr-qc/9805028.

\bibitem{horo-marolf-strc}
G.~T. Horowitz and D.~Marolf, 
% ``A new approach to string cosmology'', 
J.~High Energy Phys.\ {\bf 07}, 014 (1998). 
(gr-qc/9805207)

\bibitem{Nico} 
D.~Giulini, 
Ph.D. thesis 
(University of Cambridge, 1989).

\bibitem{topocen}  
J.~L. Friedman, 
K.~Schleich, and 
D.~M. Witt, 
Phys.\ Rev.\ Lett.\ {\bf 71}, 1486 (1993); 
Erratum, 
Phys.\ Rev.\ Lett.\ {\bf 75}, 1872 (1995). 
(gr-qc/9305017).

\bibitem{chamb-gibb}
A.~Chamblin and G.~W. Gibbons,
Phys.\ Rev.\ D {\bf 55}, 2177 (1996).
(gr-qc/9607079)

\bibitem{louko-marolf-geon}
J.~Louko and D.~Marolf, 
Phys.\ Rev.\ D {\bf 58}, 024007 (1998). 
(gr-qc/9802068) 

\bibitem{carlip-rev}
S.~Carlip,
Class.\ Quantum Grav.\ {\bf 12}, 283 (1995).
(gr-qc/9506079)

\bibitem{boersma}
J.~P. Boersma,
Phys.\ Rev.\ D {\bf 55}, 2174 (1997).

\bibitem{KASI} 
S.~Kachru and E.~Silverstein,  
Phys.\ Rev.\ Lett.\ {\bf 80}, 4855 (1998) 
(hep-th/9802183); 
A.~Lawrence, N.~Nekrasov, and C.~Vafa, 
Nucl.\ Phys.\ {\bf B533}, 199 (1998) 
(hep-th/9803015).

\bibitem{takagi-magnum}
S.~Takagi,
Prog.\ Theor.\ Phys.\ Suppl.\
{\bf 88}, 1 (1986).

\bibitem{birrell-davies}
N.~D. Birrell and P.~C.~W. Davies,
{\it Quantum Fields in Curved Space\/}
(Cambridge University Press, Cambridge, England, 1982).

\bibitem{wald-qft}
R.~M. Wald,
{\it Quantum Field Theory in Curved Spacetime and Black Hole
Thermodynamics\/}
(The University of Chicago Press, Chicago, 1994).

\bibitem{unruh-magnum}
W.~G. Unruh,
Phys.\ Rev.\ D {\bf 14}, 870 (1976).

\bibitem{israel-vacuum}
W.~Israel,
Phys.\ Lett.\ A {\bf 57}, 107 (1976).

\bibitem{Weissbluth}  
M.~Weissbluth, {\it Atoms and Molecules\/}, 
student edition 
(Academic, New York, 1978).

\bibitem{arfken}
G.~Arfken,
{\it Mathematical Methods for Physicists\/}, 
second edition
(Academic, New York, 1970).

\bibitem{terras1} 
A.~Terras, 
{\it Harmonic analysis on
symmetric spaces and applications\/}
(Springer, New York, 1985), 
Vol.~I\null. 

\bibitem{ABHP} 
S.~\AA{}minneborg, 
I.~Bengtsson, 
S.~Holst, and
P.~Peld\'an, 
Class.\ Quantum Grav.\ {\bf 13}, 2707 (1996).
(gr-qc/9604005)

\bibitem{BLP} 
D.~R. Brill, 
J.~Louko, and 
P.~Peld\'an, 
Phys.\ Rev.\ D {\bf 56}, 3600 (1997). 
(gr-qc/9705012)

\bibitem{RM}
R.~B. Mann,  
gr-qc/9709039.

\bibitem{MB} 
M.~Ba\~nados, 
Phys.\ Rev.\ D {\bf 57}, 1068 (1998). 
(gr-qc/9703040)

\bibitem{LUMA} 
M.~L\"uscher and G.~Mack, 
Commun.\ Math.\ Phys.\ {\bf 41}, 203 (1975).

\end{references}
\end{document}